%Paper: hep-ph/9306256
%From: LADINSKY@MSUPA.PA.MSU.EDU
%Date: Thu, 10 Jun 1993 10:02:57 -0400 (EDT)

%%%%%%%%%%%%%%%%%%%%%%%%%%%%%%%%%%%%%%%%%%%%%%%%%%%%%%%%%%%%%%%%%%%%%%
%%%  This is written in PHYZZX  and requires the macro tables.tex. %%%
%%%  The figures have been placed in the LANL database in uuencoded%%%
%%%  form;  there are nine pages of figures in PostScript.         %%%
%%%%%%%%%%%%%%%%%%%%%%%%%%%%%%%%%%%%%%%%%%%%%%%%%%%%%%%%%%%%%%%%%%%%%%

\input phyzzx
\input tables
\Pubnum{
 FERMILAB-Pub-93/040-T \cr
 JHU-TIPAC-930008 \cr
 MAD/PH/747 \cr
 MSUHEP 93/05 \cr
 NUB 3060-93 TH \cr
 NUHEP-TH-93-13 \cr
 UCD-93-5}
\date{June 1993}
%
%========================DEFINITIONS===================================
\overfullrule0pt
\def\ali{a_\ell^I}
\def\tli{t_\ell^I}
\def\A{{\cal A}}
\def\L{{\cal L}}
\def\V{{\cal V}}
\def\Tr{{\rm Tr\,}}
\def\Re{{\rm Re\,}}
\def\wl{W_L}
\def\wt{W_T}
\def\mh{m_H}

\def\refmark#1{[#1]}
\def\subsection#1{\par
   \ifnum\the\lastpenalty=30000\else \penalty-100\smallskip \fi
   \noindent{\enspace\it{{#1}}}\enspace \vadjust{\penalty5000}}

%
%======================\input galmac definitions=======================
\def\mynot#1{\not{}{\mskip-3.5mu}#1 }

\def\sqr#1#2{{\vcenter{\vbox{\hrule height.#2pt
        \hbox{\vrule width.#2pt height#1pt \kern#1pt
           \vrule width.#2pt}
       \hrule height.#2pt}}}}

\def\tli{t_l^I}
\def\ali{a_l^I}
\def\cite#1{\Ref~#1 }
\def\ra{\rightarrow}

%======================

\newcount\lineno
{\catcode`\`=\active \gdef`{\relax\lq}}

{\obeyspaces\global\let =\ } % let active space = control space
%==============JB's efintions
\def\ali{a_\ell^I}
\def\tli{t_\ell^I}
\def\A{{\cal A}}
\def\L{{\cal L}}
\def\V{{\cal V}}
\def\Tr{{\rm Tr\,}}
\def\Re{{\rm Re\,}}
%=================JG's definitions
\overfullrule0pt
\def\ali{a_\ell^I}
\def\tli{t_\ell^I}
\def\A{{\cal A}}
\def\L{{\cal L}}
\def\V{{\cal V}}
\def\Tr{{\rm Tr\,}}
\def\Re{{\rm Re\,}}
\def\mw{M_W}
\def\mww{M(WW)}
\def\CM{{\cal M}}
\def\wp{W^+}
\def\wm{W^-}
\def\wpm{W^{\pm}}
\def\gev{~{\rm GeV}}
\def\tev{~{\rm TeV}}

\def\fbi{~{\rm fb}^{-1}}

\def\mll{M(\ell\ell)}
\def\bmax{B_{max}}
\def\sbmin{SB_{min}}
\def\cosphill{\cos\phi_{\ell\ell}}
\def\lplm{\ell^+\ell^-}
\def\lplp{\ell^+\ell^+}

\def\anti{\overline}
\def\mt{m_t}

\def\subsection#1{\par
   \ifnum\the\lastpenalty=30000\else \penalty-100\smallskip \fi
   \noindent{\enspace\it{{#1}}}\enspace \vadjust{\penalty5000}}

\def\rta{\rightarrow}
%=========================Title Page===============================
\titlepage
%\singlespace
%\doublespace
\baselineskip 16pt
\title{\seventeenrm The Strongly Interacting $WW$ System: Gold--Plated Modes}
\vskip0.5in
\centerline{J. Bagger,${}^{(a)}$ V. Barger,${}^{(b)}$ K. Cheung,${}^{(c)}$
J. Gunion,${}^{(d)}$ T. Han,${}^{(e)}$}
\vskip10pt
\centerline{G. A. Ladinsky,${}^{(f)}$ R. Rosenfeld${}^{(g)}$ and
C.--P. Yuan${}^{(f)}$}
\vskip 0.5in
\item{(a)}{\it Department of Physics and Astronomy, The Johns Hopkins
University, \break Baltimore, MD  21218}
\item{(b)}{\it Department of Physics, University of Wisconsin,
Madison, WI  53706}
\item{(c)}{\it Department of Physics, Northwestern University,
Evanston, IL  60208}
\item{(d)}{\it Davis Institute for High Energy Physics,
Department of Physics,\break  University of California at Davis,
Davis, CA  95616}
\item{(e)}{\it Fermi National Accelerator Laboratory, P.O. Box 500, Batavia,
IL  60510}
\item{(f)}{\it Department of Physics and Astronomy, Michigan State University,
\break East Lansing, MI 48824}
\item{(g)}{\it Department of Physics, Northeastern University,
Boston, MA  02115}
\vskip0.5cm
\abstract{
In this paper we survey the signals and backgrounds for
a strongly-interacting electroweak symmetry breaking sector
at hadron supercolliders in the TeV region.
We study the process $pp \rightarrow WWX$, and
compute the rates for the ``gold-plated'' channels, where
$W^\pm \rightarrow \ell^\pm \nu$ and $Z \rightarrow \ell^+
\ell^-$ $(\ell = e,\mu)$, for a wide variety of models.
Using a forward jet-tag,
a central jet-veto and a back-to-back lepton cut to
suppress the Standard Model backgrounds,
we demonstrate that the SSC and LHC have substantial
sensitivity to strong interactions in the
electroweak symmetry breaking sector.
}\endpage

\chapter {Introduction}

\REF\lqt{
D. A. Dicus and V. S. Mathur, Phys. Rev. {\bf D7}, 3111 (1973);
B.~W.~Lee, C.~Quigg, and H.~Thacker, Phys. Rev. {\bf D16},
1519 (1977); M.~Veltman, Acta Phys. Polon. {\bf B8}, 475 (1977).}
\REF \changail{
M.~S.~Chanowitz and M.~K.~Gaillard, Nucl. Phys. {\bf B261}, 379 (1985).}
\REF\et{
J. M. Cornwall, D. N. Levin, and G. Tiktopoulos, Phys. Rev. {\bf D10},
1145 (1974); C.~Vayonakis, Lett. Nuovo Cimento {\bf 17}, 383 (1976);
G.~J.~Gounaris, R.~Kogerler, and H.~Neufeld, Phys. Rev. {\bf D34},
3257 (1986);
Y.--P. Yao and C.--P. Yuan, Phys. Rev. {\bf D38}, 2237 (1988);
J. Bagger and C. R. Schmidt, Phys. Rev. {\bf D41}, 264 (1990);
H. Veltman, Phys. Rev. {\bf D41}, 2294 (1990); H.--J. He, Y.--P. Kuang
and Xiaoyuan Li, Phys. Rev. Lett. {\bf 69}, 2619 (1992).}
\REF\effwtwo{
M. S. Chanowitz and M. K. Gaillard, Phys. Lett. {\bf B142}, 85 (1984);
G. L. Kane, W. W. Repko and W. R. Rolnick, Phys. Lett. {\bf B148}, 367 (1984);
S. Dawson, Nucl. Phys. {\bf B249}, 42 (1985).}
\REF\effwthree{J. Lindfors, Z. Phys. {\bf C28}, 427 (1985);
J.~F.~Gunion, J.~Kalinowksi, and A.~Tofighi-Niaki,
{\it Phys. Rev. Lett.} {\bf 57}, 2351 (1986);
W. B. Rolnick, Nucl. Phys. {\bf B274}, 171 (1986); P. W. Johnson, F. I. Olness
and Wu--Ki Tung, Phys. Rev. {\bf D36}, 291 (1987);
Z. Kunszt and D. E. Soper, Nucl. Phys. {\bf B296}, 253 (1988); A. Abbasabadi,
W. W. Repko, D. A. Dicus and R. Vega, Phys. Rev. {\bf D38}, 2770 (1988); S.
Dawson, Phys. Lett. {\bf B217}, 347 (1989).}

\FIG\fone{ Symbolic diagrams for the $W_LW_L \rightarrow W_LW_L$
scattering signal. The black region represents the $W_LW_L$ strongly
interacting physics.}

\FIG\ftwo{ Representative diagrams for backgrounds to the $W_LW_L$ signal:
(a) EW processes; (b) lowest-order QCD processes, with possible additional
QCD-jet radiation; and (c) top quark backgrounds.}

During the past decade, the discovery of the $W$ and the $Z$ bosons
demonstrated that the gauge structure of the Standard Model (SM)
is correct.  However, little is known about the mechanism that
gives the vector bosons their mass.  In the Standard Model, they
acquire mass because a scalar field, the Higgs doublet, has a nonzero
vacuum expectation value $v$.  At present, however, there is no
experimental evidence in favor of the Higgs particle: all the
precision measurements can be described by a Higgs-free
Standard Model.

Of course, the Standard Model without a Higgs boson cannot be a
fundamental theory \refmark{\lqt,\changail}.  It is only an effective
theory, breaking down below a few TeV.  New physics must emerge
below this scale -- which the next round of accelerators had better
be prepared to find!

$WW$ scattering provides a
particularly promising avenue for investigating this new physics
(here and henceforth $W$ generically denotes the $W$ or $Z$ boson,
unless specified otherwise).  The $WW \rightarrow WW$ cross section
without a light Higgs boson violates perturbative unitarity at about 1 TeV.
Consequently, new physics must couple to this channel in just such a way
as to cure its bad high energy behavior.

In this paper we will investigate signals and backgrounds for
the process $pp \rightarrow WWX$ at hadron supercolliders,
such as the SSC and LHC.  We will concentrate on the situation in
which there are no new particles below a TeV. We
shall study a variety of possible models, all of which are perfectly
consistent with the data to date.

Of course, in such studies one must decide what is the ``signal'' and
what is the ``background.''  We will take the {\it signal} to be the
process $pp \rightarrow W_LW_LX$, as shown in Fig.~1,
where $L$ refers to longitudinal
polarization (while the transverse polarization will be denoted by $T$).
This definition of the signal is appropriate because the $\wl\wl$
channels couple most strongly for new physics, and $\wl\wl$ production
is negligible unless the interactions among the $W$'s are strong.
Since we are mainly interested in physics for the electroweak symmetry
breaking sector, we will not include the contributions to our $\wl\wl$
signal from Yukawa couplings, such as $t \bar t H$ in the SM.
The most difficult {\it background} to the $\wl\wl$ final state is
$\wl\wt$ and $\wt\wt$ production: $pp \rightarrow W_LW_TX$ and $pp
\rightarrow W_TW_TX$.  Such processes are a background in the sense
that their cross sections are essentially independent of strong
interactions in the $W$ sector, \ie\ they are insensitive to new physics.
Further, this background is irreducible in that the final state contains
two real $W$'s analogous to the signal of interest (ignoring
polarization).

Ultimately, after appropriate cuts, the $\wl\wt+\wt\wt$
background is dominated by the ``electroweak'' (EW) diagrams, as shown
in Fig.~2a, which includes $\wt\wt,\wl\wt$ scattering diagrams and those
in which $W$'s are radiated
or emitted via electroweak interactions.  An additional contribution
to the $\wt\wl+\wt\wt$ background arises
from the $q\anti q$ annihilation processes illustrated in Fig.~2b.
Since both of these backgrounds are essentially independent of the
new physics in the $\wl\wl$ channel, we are free to compute them using
the Standard Model with a light (100 GeV) Higgs, for which $\wl\wl$
production is negligible. The difference between this computation and
a first-principles computation of the background in a model
which incorporates strong interactions in the $\wl\wl$ sector
is negligible at the energies we consider. Finally, there are heavy quark
backgrounds, especially those associated with top quark production
and decay, Fig.~2c.  These too may be reliably computed in the SM
once the top-quark mass is known.

For most of our signal estimates,
we will simplify our calculations by using the Goldstone-boson Equivalence
Theorem \refmark{\lqt-\et}, which states that, at high energies,
the external longitudinal vector bosons
can be replaced by their corresponding would-be Goldstone bosons.  This
is both a computational and conceptual simplification, for it allows us
to draw on our considerable experience with Goldstone-boson scattering
in QCD.  We will also use the effective $W$
approximation \refmark{\effwtwo,\effwthree}
to connect the $W_LW_L$ subprocesses to the $pp$ initial state.

\REF\rwwww{For a review, see C.--P. Yuan, {\it
Proposal for Studying TeV $W_L W_L
\ra W_L W_L$ Interactions Experimentally},
preprint MSUTH 92/06, 1992,
to be published in {\it Perspectives On Higgs Physics}, edited by
Gordon L. Kane, World Scientific Publishing Company, in press. }

We focus our attention on the ``gold-plated'' events, where the $W$
and $Z$ decay to charged leptonic final states $(\ell = e,\mu)$.
For the purpose of this study,
we ignore final states where the bosons decay hadronically, as well as final
states where either of the $Z$'s decays into neutrinos.  These final
states should also be studied and will possibly improve
the observability of electroweak symmetry breaking at
the SSC/LHC \refmark{\rwwww}.

Because we focus on the gold-plated leptonic channels, the only
backgrounds to the $\wl\wl$ signal that we need to consider are those
in which real $\wl\wt$ and $\wt\wt$ pairs are produced.
As already noted, in the final analysis, the diagrams in Fig.~2a
yield the most difficult backgrounds.
We suppress these backgrounds by imposing further restrictions
on the events. However, we must also deal with the additional
background processes of Figs.~2b and 2c.
The continuum pair production processes of Fig.~2b
arising from $q\anti q$ annihilation (which we term the QCD background)
contribute to the $W^+W^-$, $W^\pm Z$ and $ZZ$ channels.
At lowest order, these annihilation processes have a very different final state
structure than the $WW$ scattering processes of interest,
where spectator quark jets are left behind when the incoming
quarks radiate the initial-state $W$'s that
then scatter (see Fig.~1).  Thus, even allowing for higher-order radiation
corrections, the QCD background can be greatly suppressed
by requiring a tagged forward jet.
The heavy top quark processes of Fig. 2c,
arising from $t\anti t$, $t\anti t W$ and $t\anti t Z$  production
followed by $t$ and $\anti t$ decays to real $W$'s,
contribute to the $W^+W^-$, $W^\pm Z$ and $W^\pm W^\pm$ channels.
Fortunately, these top quark background
processes have substantial jet activity at moderate rapidity and
can be efficiently suppressed by requiring a central jet-veto.

Indeed, it turns out that both the forward jet-tag and the central jet-veto are
effective in reducing the backgrounds from the irreducible $LT+TT$
electroweak backgrounds as well. Nonetheless, if only jet-tagging
and/or vetoing is applied, a substantial EW $LT+TT$ background remains
in the $\wpm\wpm$ and $\wp\wm$ channels. This background remnant
can be greatly reduced with little impact on the $LL$
signal by requiring energetic leptons at low rapidity, and especially
requiring that the two leptons appearing in the final
state be very back-to-back.

Because we use the effective $W$ approximation for our signal, we can only
estimate the effects of the tag and veto cuts.  We use the exact
Standard Model calculation with a 1 TeV Higgs to derive efficiencies for
these cuts. Since these efficiencies should be relatively model-independent,
we can apply them to the effective $W$ calculations to estimate the rate
for each signal. The efficiency for the lepton cuts,
including the back-to-back requirements if imposed, is obtained
by employing the effective $W$ approximation and decaying the final
$W$'s appropriately. The accuracy of this procedure was tested in
the SM 1 TeV Higgs case.  Good agreement was found between the lepton
cut efficiencies obtained in the exact
calculation and in the effective $W$ calculation.

In section~2 of this paper we study these procedures for the Standard Model.
We take the signal to be a 1 TeV Higgs resonance, and the
electroweak background to be the SM rate for a light Higgs boson
(we employ $\mh=100$ GeV). We present the cuts that maximize
the signal/background ratio while preserving a reasonable rate.  We use the
exact calculation to compute efficiencies for the forward jet-tag and the
central jet-veto. In section~3 we present the different models we will employ.
We examine resonant and nonresonant scenarios, and frame our discussion in
the language of chiral lagrangians.  In section~4 we examine the
accuracy of our procedure in which we apply the cut efficiencies obtained
from the exact SM calculation of section~2 to the cross sections obtained
using the equivalence theorem and the effective $W$ approximation.
We then present our
basic numerical results and assess the reach of the SSC and the LHC for
each of the strongly interacting $W$-system models.
Section~5 contains further discussion and comments.
We conclude with some brief remarks in section~6.

\chapter {Standard Model}

\REF\trivial{J. Kuti, L. Lin, and Y. Shen, Phys. Rev. Lett. {\bf 61}, 678
(1988); M. Luscher and P. Weisz, Phys. Lett. {\bf B212}, 472 (1988);
Nucl. Phys. {\bf B318}, 705 (1989); U. M. Heller {\it et al.},
FSU-SCRI-93-29 (1993).}
\REF\gunglue{J.~F.~Gunion \etal, Ref. \effwthree.}
\REF\jtone{R.~N.~Cahn {\it et al.}, Phys.\ Rev.\ {\bf D35}, 1626 (1987);
V.~Barger, T.~Han, and R.~J.~N.~Phillips, Phys.\ Rev.\ {\bf D37} 2005 (1988);
R.~Kleiss and W.~J.~Stirling, Phys.\ Lett.\ {\bf 200B}, 193 (1988).}
\REF\jttwo{D.~Froideveaux, Proceedings of the LHC Workshop (1990),
CERN 90-10, Vol~II, p.~444; M.~H.~Seymour, {\it ibid}, p.~557.}
\REF\jtthree{
U.~Baur and E.~W.~N.~Glover, Nucl.\ Phys.\ {\bf B347}, 12 (1990);
U.~Baur and E.~W.~N.~Glover, Phys.\ Lett.\. {\bf B252}, 683 (1990).}
\REF\wpwpb{V.~Barger, K.~Cheung, T.~Han, and R.J.N.~Phillips,
 Phys.~Rev. {\bf D42}, 3052 (1990).}
\REF\hzz{V.~Barger, K.~Cheung, T.~Han, J.~Ohnemus, and D.~Zeppenfeld,
 Phys.\ Rev.\ {\bf D44}, 1426 (1991).}
\REF\hww{V.~Barger, K.~Cheung, T.~Han, and D.~Zeppenfeld,
Phys.\ Rev.\ {\bf D44}, 2701 (1991); Erratum to appear in
Phys.\ Rev.\ {\bf D}; MAD/PH/757.}
\REF\hwz{V.~Barger, K.~Cheung, T.~Han,  A.~Stange, and D.~Zeppenfeld,
Phys.\ Rev.\ {\bf D46}, 2028 (1992).}
\REF\wpwpg{D.~Dicus, J.F.~Gunion, and R.~Vega, Phys.~Lett.~{\bf 258B}, 475
(1991).}
\REF\wpwpo{D.~Dicus, J.F.~Gunion, L.~H.~Orr, and R.~Vega, Nucl.\ Phys.\  {\bf
B377}, 31 (1991).}
\REF\isol{talks given by W.~Smith,  F.~Paige, and T.~Han
at the {\it SSC Physics Symposium},
Madison, Wisconsin, Feb.~1991 (unpublished).}
\REF\zz{R.~W.~Brown and K.~O.~Mikaelian, Phys.\ Rev.\ {\bf D19}, 922 (1979).}
\REF\zzjets{ U.~Baur, E.~W.~N.~Glover, and J.~J.~van~der~Bij, Nucl.~
Phys.~{\bf B318}, 106 (1989);
V.~Barger, T.~Han, J.~Ohnemus, and D.~Zeppenfeld,
Phys.~Rev.~{\bf D41}, 2782 (1990).}
\REF\zzloop{J.~Ohnemus and J.~F.~Owens, Phys.~Rev.~{\bf D43}, 3626 (1991);
B.~Mele, P.~Nason, and G.~Ridolfi, Nucl.\ Phys.\ {\bf B357}, 409 (1991). }
\REF\ggzz{D.~A.~Dicus, C.~Kao, and W.~W.~Repko, Phys.\ Rev.\
{\bf D36}, 1570 (1987); E.~W.~N.~Glover and J.~J.~van~der~Bij, Nucl.\ Phys.\
{\bf B321}, 561 (1989).}
\REF\zzqq{D.~A.~Dicus, S.~L.~Wilson, and R.~Vega, Phys.\ Lett.\
{\bf 192B}, 231 (1987).}
\REF\ptz{V.~Barger, T.~Han, and R.~J.~N.~Phillips,
Phys.\ Lett.\ {\bf 200B}, 193 (1988); U.~Baur and E.~W.~N.~Glover,
Phys.\ Rev.\ {\bf D44}, 99 (1991). }
\REF\tttree{R.~K.~Ellis and J.~C.~Sexton, Nucl.\ Phys.\ {\bf B282},
642 (1987).}
\REF\ttloop{P.~Nason, S.~Dawson, and R.~K.~Ellis, Nucl.\ Phys.\ {\bf B303},
607 (1988); W.~Beenakker {\it et al.}, Phys.\ Rev.\ {\bf D40}, 54 (1989);
Nucl.\ Phys.\ {\bf B351}, 507 (1991).}
\REF\ttdecay{R.~Kleiss and W.~J.~Stirling, Z.~Phys. {\bf C40}, 419 (1988);
H.~Baer, V.~Barger, and R.~J.~N.~Phillips, Phys.\ Rev.\ {\bf D39},
2809 (1989); H.~Baer, V.~Barger, J.~Ohnemus, and R.~J.~N.~Phillips,
Phys.\ Rev.\ {\bf D42}, 54 (1990);
R.~P.~Kauffman and C.P.~Yuan, Phys.~Rev.~ {\bf D42}, 956 (1990);
G.~A.~Ladinsky and C.P.~Yuan, Phys.~Rev.~ {\bf D43}, 789 (1990). }
\REF\mtb{V.~Barger, A.~D.~Martin, and R.~J.~N.~Phillips, Phys.\
Lett.\ {\bf B125}, 339(1983); V.~Barger, T.~Han, and J.~Ohnemus, Phys.\ Rev.\
{\bf D37}, 1174(1988).}
\REF\sdc{SDC Technical Design Report, SDC-92-201 (1992).}
\REF\wpwp{M.~S.~Chanowitz and M.~Golden, Phys.\ Rev.\ Lett.\ {\bf 61}, 1053
(1988); {\bf 63}, 466(E) (1989).}
\REF\wpwpbc{M.~S.~Berger and M.~S.~Chanowitz, Phys.~Lett.~{\bf B263},
509 (1991).}
\REF\gwpwp{D.~A.~Dicus and R.~Vega, Phys.~Lett.~{\bf B217}, 194 (1989).}

In this section, we discuss $W W$ scattering in the Standard Model
with a 1~TeV Higgs particle.  Although it is argued \refmark{\trivial}
that the SM is not a consistent effective theory if $m_H \gsim 800$~GeV
or so, we take this case as a prototype for models with strong $WW$
scattering.
We present the signal and the background,
calculated using the exact, order $\alpha^2$, matrix elements for
$pp \rightarrow W W X$.  We use these results to derive efficiencies
for the forward jet-tag and the central jet-veto.
The comparison of signal results to those found using the equivalence theorem
and the effective $W$ approximation will be presented in sections~4 and~5.

In this study we concentrate on the purely leptonic decay modes of the
final state $W$'s, namely the ``gold--plated'' events,
with $W^\pm \rightarrow \ell^\pm \nu_\ell$ and
$Z \rightarrow \ell^+\ell^-$ $(\ell = e,\mu)$.  The experimental signature is
given by two or more isolated, charged leptons in the central rapidity
$(y(\ell))$ region, with large transverse momenta $(p_T)$.
Although clean, these gold-plated channels carry the price of
relatively small branching fractions for the purely leptonic $W$ decays.

The diagram for longitudinal vector boson scattering is given symbolically
in Fig.~1,
$$
qq\ \rightarrow\ qqW_LW_L\ ,
\eqn\twoone
$$
where $W_L$ denotes a longitudinally-polarized vector boson $(W_L = W_L^\pm,\,
 Z_L)$.
If the interactions between $W_L$ bosons are strong at high energies,
we expect $W_LW_L$ scattering to be enhanced at large invariant mass.
It is this enhancement which defines the signal we wish to isolate.

The irreducible backgrounds are shown in Fig. 2.  At least one of
the final $W$'s produced in the background processes is transversely
polarized.  In particular, the cross sections for
$WW$ scattering to produce $\wl\wt$ or $\wt\wt$
pairs are essentially independent of the Higgs mass in the SM and are
part of the background by definition.
Other backgrounds include gluon-exchange between quarks with initial and
final state emission of two $W$'s (both of which are dominantly transversely
polarized) \refmark{\gunglue}, and a variety of electroweak processes in which
a final $W_T$ arises via bremsstrahlung
or emission from a primary quark or electroweak boson line.
Continuum $WW$ pair production arising from $q\anti q$ annihilation and
$gg$ fusion also contributes to the background.  For cases with a $W^\pm$ in
the
final state, there is an especially
important {\it reducible} background from heavy quark production and decay.

It is important to note that two spectator quarks always emerge in
association with the $\wl\wl$ scattering signal, but that spectators emerge
in only a subset of the irreducible backgrounds.
The spectator quarks usually appear in forward/backward regions, and have
an energy of order one TeV and a $p_T$ of order $M_W/2$. It is therefore
possible to improve the signal/background ratio by tagging those
quark jets (in particular, continuum pair production processes do not
have a spectator quark jet at lowest order in perturbation
theory) \refmark{\jtone}.
While studies have shown that tagging
{\it two} high $p_T$ spectator jets
substantially enhances the signal/background ratio,
such double tagging proves to be too costly
to the signal \refmark{\jttwo-\hzz}.
It has been recently suggested that
tagging just one of these quarks as a single {\it energetic} jet
can be just as efficient in suppressing the backgrounds
that do not intrinsically require spectator jets,
and far more efficient in retaining the signal for a
heavy Higgs boson \refmark{\hzz-\hwz}.
Thus, to isolate the heavy Higgs and other types of strong $W_LW_L$
signals, we will apply such a forward jet-tag for most final state
channels \refmark{\hzz-\wpwpo}.

The detailed characteristics of $\wl$ emission and the associated
spectator jets also play a role in separating $\wl\wl$ scattering from the
background processes which {\it do} yield spectator jets (as well as two
$W$'s) in the final state.
The crucial point to note is that the initial $\wl$'s
participating in the $\wl\wl$ scattering have a $1/(p_T^2+M_W^2)^2$
distribution with respect to the quarks from which they are emitted.
This is to be contrasted, for instance, with $\wt\wt$ scattering where
the initiating $\wt$'s have a $p_T^2/(p_T^2+M_W^2)^2$ distribution with
respect to the emitting quarks.
The softer $p_T$ distribution in the $\wl\wl$ case has two
primary consequences.  First, the final $\wl\wl$ pair is likely to have
much more limited net transverse motion than $\wl\wt$ and $\wt\wt$
pairs produced through the various irreducible backgrounds.
Secondly, the spectator quarks left
behind tend to emerge with smaller $p_T$ (order of $M_W/2$), and
correspondingly larger
rapidity, than those associated with the background processes
containing spectator jets and $\wl\wt$ or $\wt\wt$ pairs.

There are several crucial secondary consequences resulting from the above
special characteristics of $\wl\wl$ scattering.
First, as discussed above, the jets from the gluon-exchange background and the
electroweak background are generally harder and more central than those from
the signal \refmark{\wpwpb,\wpwpg}.
Therefore we will normally veto hard central jets to enhance the
signal/background ratio \refmark{\wpwpb,\wpwpg,\wpwpo}. Such a veto retains
most of the signal events. As a further bonus,
a central jet-veto is especially effective in suppressing
the reducible background from heavy quark production and decay.
The jets associated with this latter type of background populate a
much more central region than do those from spectator quarks.
Another consequence of the small $p_T$ of the $\wl\wl$ system is that
we expect the charged leptons from the decays of the two final
$\wl$'s to be very back-to-back in the transverse plane
\refmark{\wpwpg,\wpwpo}.  This
is due not only to the limited $p_T$ of the $\wl\wl$ system but also to the
fact that the bulk of the leptons emitted from each final $\wl$
will have a significant (and relatively similar) fraction of the $\wl$'s total
momentum. The latter fact also implies that the leptons will generally be
very energetic.
A cut requiring that the leptons
appearing in the final state be very energetic and
very back-to-back will substantially
reduce all backgrounds, while being highly efficient in retaining
the $\wl\wl$ signal events.

We have already noted that the charged leptons will be required to be isolated.
In order to completely eliminate the background from heavy quark production
and decay (say, $b$ or $c$ semileptonic decays)
in all channels, we implicitly assume that it will be possible
to implement an isolation requirement according to which the hadronic
energy deposit within a cone ${\Delta R} < 0.3$ around the lepton
must be less than about 5 GeV \refmark{\isol}.

Before proceeding, we wish to reemphasize the precise definition
of the signal and background that we employ.  The
results in this section will all be based on the full matrix element
calculations for the Standard Model.  We define the heavy
Higgs boson signal to be the difference between the cross section with
a heavy Higgs boson and the result with a light Higgs; for example,
$$
\sigma({\rm signal~for~a~1~TeV~Higgs~boson})\ = \
\sigma(m_H=1~{\rm TeV}) - \sigma(m_H=100~{\rm GeV})\ ,
\eqn\twothirteen
$$
where all $W$ helicities have been included for both $m_H$ values.
At SSC energies, the EW rate for production of $W$ pairs in which
one or both of the $W$'s is transversely polarized is essentially
independent of the Higgs mass, while $\wl\wl$ production is
extremely small at $m_H=100~{\rm GeV}$.  Thus,
the prescription \twothirteen\ measures the production rate of
longitudinally polarized $W$ bosons at large $\mh$. We will sometimes refer
to this definition of the signal as the ``subtraction'' result.

As stated earlier in the introduction, we are
only interested in the EW symmetry breaking sector; we
do not include contributions to the $\wl\wl$
final state arising from processes such as
$gg, q\bar q \ \rightarrow \ t \bar t H$ and $gg \ \rightarrow \ H$ (via
a top quark loop) that depend upon the Yukawa couplings of the Higgs boson.

We now turn to a detailed discussion of the signals and backgrounds for
the leptonic decay modes associated with each of the
possible $W_LW_L$ scattering channels.

1) $W^+W^-\ \rightarrow\ ZZ$, $ZZ\ \rightarrow\ ZZ$

We first consider the ``gold-plated'' events with four charged leptons
from $ZZ$ decays.  This gives a clean and distinct signal because
the $ZZ$ pairs can be fully reconstructed.  The disadvantage
is the rather small leptonic branching fraction, BR$(ZZ \rightarrow
4\ell)=0.45\%$.

The major Standard Model backgrounds for this process
arise from continuum $ZZ$ production via tree-level processes
at ${\cal O}(\alpha^2)$, ${\cal O}(\alpha^2\alpha_s)$ and
${\cal O}(\alpha^2\alpha_s^2)$ \refmark{\zz-\zzloop}; for example,
$$
q q\ \rightarrow\ ZZ\ +\ {\rm jets}\ ,
\eqn\twothree
$$
which we will refer to as the QCD background. This set of backgrounds
includes, in particular, diagrams at order $\alpha^2\alpha_s^2$ from
gluon-exchange diagrams
in which two quarks scatter via gluon exchange while two vector bosons are
emitted from either the initial or final quark lines. Since the
$Z$'s are
relatively weakly coupled to quark lines, this type of background is
small in this case.
At supercollider energies, the one-loop process
$$
gg\ \rightarrow\ ZZ
\eqn\twofour
$$
is also not negligible. The total production rate is 30-70\% as large as
that from $q\bar q\rta ZZ$, depending on the top quark mass
\refmark{\ggzz}.  However, we are interested in the large $M(ZZ)$
region, and require a very energetic jet in the final state, so the
effective gluon luminosity is suppressed to a level where we can ignore
gluon fusion in our calculations.

The ${\cal O}(\alpha^4)$
electroweak production of transversely polarized $Z$-pairs
is another irreducible background
\refmark{\jtthree,\hzz,\zzqq}.  Although it is formally higher-order than
Eq. \twothree\ in terms of the electroweak coupling constant, the kinematics
are so similar to the signal that it must also be included.
We will refer to this as the electroweak (EW) background.

In a recent study, kinematical cuts were developed to suppress these
backgrounds for detecting a heavy SM
Higgs boson at the SSC and the LHC \refmark{\hzz}.
We will use the same cuts,
$$\eqalign{
p_T^{}(\ell)\ >&\ 40\ {\rm GeV}\ ,\ \qquad |y(\ell)|\ <\ 2.5\ , \cr
M(ZZ)\ >&\ 500\ {\rm GeV}\ , \qquad p_T(Z)\ > \
{1 \over 4} \sqrt{M(ZZ)^2 - 4M_Z^2}\ , \cr}
\eqn\twofive
$$
where $y(\ell)$ is the rapidity of the lepton $\ell$, and $M(ZZ)$ is the
invariant mass of the two $Z$'s in the final state.  The transverse momentum
cut on the $Z$'s is motivated by its facility in removing the QCD background
\refmark{\ptz}.
As discussed above, a forward (or backward) jet-tag is
very effective in suppressing the QCD and EW backgrounds \refmark{\hzz}.
Therefore, we will also require a tagged jet in the region
$$
E(j_{\rm tag})\ >\ 1.0\ (0.8)\ {\rm TeV}\ , \qquad 3\ <\
|y(j_{\rm tag})|\ <\ 5\
,
\qquad
p_T(j_{\rm tag})\ >\ 40\ {\rm GeV}\ ,
\eqn\twosix
$$
where the number outside (inside) the parentheses refers to the
cut applied at the SSC (LHC).

The jet-tagging efficiency is about 60\% for the signal.  The combined
cuts essentially eliminate the QCD background and substantially
suppress the EW background. An additional cut requiring the leptons
from opposite $Z$'s to be back-to-back is not needed in this case.

2) $W^+W^-\ \rightarrow\  W^+W^-$, $ZZ\ \rightarrow\  W^+W^-$

We next consider $W^+W^-$ events in the $\ell \bar\nu_\ell \bar \ell
\nu_\ell$ final state, where $\ell=e,\mu$. The leptonic branching fraction
is BR$(WW \rightarrow \ell \bar\nu_\ell \bar\ell \nu_\ell)=4.7 \%$,
so  we expect a larger number of events in this channel. Although the two
$W$'s cannot be fully reconstructed, any $s$-channel resonance, such as the
Standard Model Higgs boson, significantly enhances the production rate, and
the $M(\ell\ell)$ spectrum peaks broadly at about one-half the
resonance mass \refmark{\hww}.

Unfortunately, there are now reducible backgrounds besides the irreducible
backgrounds from continuum QCD and EW processes.
The most important ones are
$$
q \bar q, gg \rightarrow t \bar t\ ,
\quad gg \rightarrow t \bar t g\ ,
\quad qg \rightarrow t \bar t q\ ,
\quad q \bar q \rightarrow t \bar t g\ ,
\eqn\twoseven
$$
where the top quarks decay into real $W$'s \refmark{\tttree-\ttdecay}.

To reduce the backgrounds, we first impose stringent leptonic
cuts
$$\eqalign{
p_T(\ell)\ >\ & 100\ {\rm GeV}\ , \qquad |y(\ell)|\ <\ 2\ , \cr
\Delta p_T(\ell\ell)\ \equiv\ & |{\bf p_T}(\ell_1) - {\bf p_T}(\ell_2)|
\ > \ 450\ {\rm GeV}\ , \cr
M(\ell\ell) >\ & 250\ {\rm GeV}\ , \qquad {\rm cos}\phi_{\ell\ell}\ < \  -0.8\
, \cr}
\eqn\twoeight
$$
where $\Delta p_T(\ell\ell)$ and cos$\phi_{\ell\ell}$ are,
respectively, the difference of the transverse momenta and the cosine of
the opening angle in the transverse plane of the two charged leptons.
The cuts on these two variables are based on our earlier observations
that the lepton-pair decay products are more energetic
and more back-to-back in the transverse plane for signal events than for the
backgrounds \refmark{\wpwpg}. For example, the transverse momentum
of the charged leptons, $p_T(\ell)$, for the signal will
typically be of order $m_H/4$. For a 1 TeV Higgs
boson, $\Delta p_T(\ell\ell)\ \sim m_H/2 = 500$~GeV.

We also impose the jet-tagging conditions \refmark{\hww}
$$
E(j_{\rm tag})\ >\ 1.5\ (1.0)\ {\rm TeV}\ ,\qquad 3\ <\ |y(j_{\rm tag})|
\ <\ 5\ , \qquad
p_T(j_{\rm tag})\ >\ 40\ {\rm GeV}\ .
\eqn\twonine
$$
The $E(j_{\rm tag})$ cut has been made slightly more stringent
than Eq.~\twosix\ in order to control the much larger
$t \bar t g$ background.   We further suppress the top
background by a central jet-veto in which events with jets with \refmark{\hww}
$$
p_T(j_{\rm veto})\ >\ 30\ {\rm GeV}\ ,\qquad |y(j_{\rm veto})|\ < 3\
\eqn\twoten
$$
are rejected. In Ref.~{\sdc}, a central jet threshold of 25 GeV was used
by the SDC collaboration. Our choice in Eq.~\twoten~is slightly more
conservative.

Combining the cuts of Eqs.~\twoeight-\twoten, we can reduce the
backgrounds below the $\wl^+\wl^-$ signal. Especially significant is the
effective reduction of the large $t\bar t$ background.
With the leptonic cuts of Eq.~\twoeight\ imposed, the overall
efficiency for jet-tagging and vetoing is about 38\% for the signal.
We have chosen $m_t=140$ GeV as representative in our background analyses
throughout this paper. If the top quark is heavier, our jet-veto cut
would be more effective and the $t \bar t j$ background would be easier to
separate \refmark{\hww}.

3) $W^+Z\ \rightarrow\ W^+Z$

We now turn to the $WZ$ events with $\ell \bar\nu_\ell \ell \bar\ell $
final states.  The leptonic branching fraction is
BR$(W^+Z \rightarrow \ell  \bar\nu_\ell \ell \bar\ell )=1.5 \%$.

For this channel, we choose the leptonic acceptance cuts as follows,
$$\eqalign{
p_{T}(\ell )\ >&\ 40\ {\rm GeV}\ , \qquad |y(\ell)|\ <\ 2.5\ , \cr
{\mynot p}_{T}\ >&\ 75\ {\rm GeV}\ ,\qquad M_T\ >\  500\ {\rm GeV}\ ,
\ \ p_T(Z)\ >\ {1 \over 4}M_T\ , \cr}
\eqn\twoeleven
$$
where ${\mynot p}_{T}$ denotes the missing transverse momentum and
$M_T$ is the cluster transverse mass of the $WZ$ system, defined
by \refmark{\mtb}
$$
M_T^2\ =\ \bigg(\sqrt{M^2(\ell \ell \ell) + p_T^2(\ell \ell \ell)}
+ |{\mynot p}_T|\bigg)^2 -
( {\bf p}_T(\ell \ell \ell) + {\bf {\mynot p}}_T )^2\ .
\eqn\twotwelve
$$
Analogous to the cut on $p_T(Z)$ given in Eq.~\twofive,
the $p_T(Z)$ cut in Eq.~\twoeleven\ is useful for removing the QCD background.

To reduce the QCD and EW backgrounds, following Ref.~\hwz,
we tag a jet with
$$
E(j_{\rm tag})\ >\ 2.0\ (1.5)\ {\rm TeV}\ ,\qquad 3\ <\ |y(j_{\rm tag})|
\ <\ 5\ ,
\ \
p_T(j_{\rm tag})\ >\ 40\ {\rm GeV}\ .
\eqn\twonine
$$
We can reduce the background from $Z$ and top quark associated production,
$$
q \bar q, gg\ \rightarrow\ Z t \bar t\ ,
\eqn\twothirteenn
$$
by imposing the jet-vetoing of Eq.~\twoten.

The tagging plus vetoing efficiency is about 40\% for the signal.
As for the fully reconstructable $ZZ\rightarrow 4\ell$ mode, a back-to-back
lepton cut is not needed here.

4) $W^+W^+\ \rightarrow\ W^+W^+$

Finally, we discuss the like-sign $W$ process with two like-sign charged
leptons in the final state \refmark{\wpwp,\wpwpb,\wpwpg,\wpwpbc,\wpwpo}.
This mode is attractive
because of the distinctive final state and absence of an order $\alpha^2$
continuum background.

However, backgrounds to the $W^+_LW^+_L$ signal do exist.
Besides the transversely polarized background from EW processes,
there is the previously-mentioned background contributing
at order $\alpha^2\alpha_s^2$,
$$
q q \rightarrow\ q q W^+W^+\ ,
\eqn\twofourteen
$$
in which a gluon is exchanged between the scattering quarks
\refmark{\gunglue,\gwpwp}.
Since there is no lowest order ($\alpha^2$ or $\alpha^2\alpha_s$) background,
this process is potentially significant for this channel.
Finally, there is a background from associated $Wt \bar t$
production \refmark{\wpwpb},
$$
q' \bar q \ \rightarrow\ Wt \bar t \ ,\eqn\twothirteennn
$$
with $t \rightarrow W^+b$.

We first impose the leptonic cuts of Eq.~\twoeight, with the exception
of a weaker cut $\Delta p_T(\ell\ell)\ > 200$~GeV.
The back-to-back cuts are advantageous in the present channel
\refmark{\wpwpg,\wpwpo}.
We also apply the jet-vetoing of Eq.~\twoten\ to this case,
with a looser cut $p_T(j_{\rm veto})\ >\ 60$ GeV
\refmark{\wpwpb}, and find that it greatly reduces the backgrounds.

Another potentially large background is that from $t\bar t$ production with
a cascade decay, $\bar t \rightarrow \bar b W^- \rightarrow \ell^+ X $.
However, the $\ell^+$ from the $\bar b$ decay is usually not isolated.
When the $\ell^+$ is fast, the other hadrons from the $\bar b$
decay tend to be collinear with the $\ell^+$.  The
lepton isolation requirement that we have implicitly assumed should be
able to eliminate this cascade decay background \refmark{\wpwpo,\isol}.

With the leptonic cuts imposed,
the jet-vetoing efficiency for this signal is about 70\%.

Jet-tagging can also be applied to the $W^+W^+$ process
\refmark{\wpwpg,\wpwpo}.
By tagging a forward jet and imposing a cut on the minimum invariant
mass of the tagged jet and a lepton, $M(\ell j_{\rm tag})>200$ GeV, it is
possible to further reduce the backgrounds.  This tag is especially
effective to reduce the $t\bar t$ cascade decay background
since $M(\ell j_{\rm tag})$
is significantly larger for the signal than for the background.
However, since we assume that charged lepton isolation can be implemented
at the level required to eliminate the cascade decay background,
we will not impose such a cut in this paper.  Should a problem
arise in experimentally implementing lepton isolation, this type
of cut can be used as an alternative.

\TABLE\cuttable{}
\topinsert
\titlestyle{\twelvepoint
Table~\cuttable : Leptonic cuts, tagging and vetoing cuts on jets,
by mode at the SSC (LHC).}
\bigskip
\thicksize=0pt
\hrule \vskip .04in \hrule
\begintable
$Z Z$ leptonic cuts & Tag only \cr
 $\vert  y({\ell}) \vert  < 2.5 $  &
 $E(j_{tag}) > 1.0\ (0.8)\ {\rm TeV}$   \nr
 $p_T(\ell) > 40\ {\rm GeV}$  &
 $3.0 < \vert y(j_{tag}) \vert < 5.0$   \nr
 $p_T(Z) > {1\over4} \sqrt{M^2({ZZ}) - 4 M^2_Z}$  &
 $p_T(j_{tag}) > 40\ {\rm GeV}$   \nr
 $M({ZZ}) > 500\ {\rm GeV}$  & {} \cr
$W^+ W^-$ leptonic cuts & Tag and Veto \cr
 $\vert y({\ell}) \vert < 2.0 $  &
 $E(j_{tag}) > 1.5\ (1.0)\ {\rm TeV}$  \nr
 $p_T(\ell) > 100\ {\rm GeV}$  &
 $3.0 < \vert y(j_{tag}) \vert < 5.0$  \nr
 $\Delta p_T({\ell\ell}) > 450\ {\rm GeV}$  &
 $p_T(j_{tag}) > 40\ {\rm GeV}$  \nr
 $\cos\phi_{\ell\ell} < -0.8$  &
 $p_T(j_{veto}) > 30\ {\rm GeV}$  \nr
 $M({\ell\ell}) > 250\ {\rm GeV}$  &
 $ \vert  y(j_{veto}) \vert  < 3.0$  \cr
$W^+ Z$ leptonic cuts & Tag and Veto\cr
 $\vert  y({\ell}) \vert  < 2.5 $  &
 $E(j_{tag}) > 2.0\ (1.5)\ {\rm TeV}$  \nr
 $p_T(\ell) > 40\ {\rm GeV}$  &
 $3.0 < \vert y(j_{tag}) \vert < 5.0$  \nr
 $ {\mynot p}_{T} >  75\ {\rm GeV}$  &
 $p_T(j_{tag}) > 40\ {\rm GeV}$   \nr
 $p_T(Z) > {1\over4} M_T $ &
 $p_T(j_{veto}) > 60\ {\rm GeV}$  \nr
 $M_T > 500\ {\rm GeV}$ &
 $ \vert  y(j_{veto}) \vert  < 3.0$  \cr
$W^+ W^+$ leptonic cuts & Veto only \cr
 $\vert  y({\ell}) \vert  < 2.0 $   &
 $p_T(j_{veto}) > 60\ {\rm GeV}$  \nr
 $p_T(\ell) > 100\ {\rm GeV}$   &
 $ \vert  y(j_{veto}) \vert  < 3.0$  \nr
 $\Delta p_T(\ell\ell) > 200\ {\rm GeV}$   & \nr
 $\cos\phi_{\ell\ell} < -0.8$  & \nr
 $M({\ell\ell}) > 250\ {\rm GeV}$ & {}  \endtable
\hrule \vskip .04in \hrule
%{$^*$  $M_T$ is the cluster transverse mass.$^5$}
%
\endinsert

\TABLE\smssctable{}
\topinsert
\titlestyle{\twelvepoint
Table~\smssctable a: Standard Model cross sections (in fb)
for $m_H=1$ TeV, $m_H=0.1$ TeV,  and for the QCD
background, with $\sqrt{s} = 40$ TeV,  $m_t = 140$ GeV.}
\bigskip
\thicksize=0pt
\hrule \vskip .04in \hrule
\begintable
$Z Z $ & leptonic cuts only & tag only & veto plus tag \cr
$ EW(m_H = 1~TeV)  $ & 1.2  & 0.68 & - \nr
$ EW(m_H = 0.1~TeV)$ & 0.17 & 0.07 & - \nr
$ QCD              $ & 0.92 & 0.02 & - \cr
$W^+ W^-$ & leptonic cuts only & veto only & veto plus tag \cr
$ EW(m_H = 1~TeV)  $ & 11  & 5.5  & 3.6  \nr
$ EW(m_H = 0.1~TeV)$ & 2.2 & 0.49 & 0.30 \nr
$ QCD              $ & 15  & 15   & 0.31 \nr
$  t \bar t j      $ & 1300& 14   & 1.5  \cr
$W^+ Z$ & leptonic cuts only & veto only & veto plus tag \cr
$ EW(m_H = 1~TeV)  $ & 1.3 & 0.41 & 0.21 \nr
$ EW(m_H = 0.1~TeV)$ & 1.1 & 0.26 & 0.13 \nr
$ QCD              $ & 3.1 & 3.1  & 0.11\nr
$  Z t \bar t j    $ & 1.4 & 0.04 & 0.01 \cr
$W^+ W^+$ & leptonic cuts only & veto only & veto plus tag \cr
$ EW(m_H = 1~TeV)  $ & 2.4  & 0.98 & - \nr
$ EW(m_H = 0.1~TeV)$ & 1.4  & 0.29 & - \nr
$ QCD              $ & 0.24 & 0.01 & - \nr
$  W t \bar t      $ & 0.75 & 0.05 & - \endtable

\hrule \vskip .04in \hrule
\endinsert

%\TABLE\smlhctable{}
\topinsert
\titlestyle{\twelvepoint
Table~\smssctable b: Standard Model cross sections (in fb)
for $m_H=1$ TeV, $m_H=0.1$ TeV,  and for the QCD
background, with $\sqrt{s} = 16$ TeV,  $m_t = 140$ GeV.}
\bigskip
\thicksize=0pt
\hrule \vskip .04in \hrule
\begintable
$Z Z $ & leptonic cuts only & tag only & veto plus tag \cr
$ EW(m_H = 1~TeV)  $ & 0.17  & 0.076 & - \nr
$ EW(m_H = 0.1~TeV)$ & 0.029 & 0.007 & - \nr
$ QCD              $ & 0.33  & 0.003 & - \cr
$W^+ W^-$ & leptonic cuts only & veto only & veto plus tag \cr
$ EW(m_H = 1~TeV)  $ & 1.6  & 0.52  & 0.31  \nr
$ EW(m_H = 0.1~TeV)$ & 0.42 & 0.049 & 0.022 \nr
$ QCD              $ & 4.3  & 4.3   & 0.042 \nr
$  t \bar t j      $ & 107  & 1.5   & 0.12  \cr
$W^+ Z$ & leptonic cuts only & veto only & veto plus tag \cr
$ EW(m_H = 1~TeV)  $ & 0.25 & 0.059 & 0.022 \nr
$ EW(m_H = 0.1~TeV)$ & 0.20 & 0.035 & 0.012 \nr
$ QCD              $ & 1.2 & 1.2  & 0.011\nr
$  Z t \bar t j    $ & 0.085 & 0.003 & 0.000 \cr
$W^+ W^+$ & leptonic cuts only & veto only & veto plus tag \cr
$ EW(m_H = 1~TeV)  $ & 0.43  & 0.13  & - \nr
$ EW(m_H = 0.1~TeV)$ & 0.27  & 0.037 & - \nr
$ QCD              $ & 0.063 & 0.003 & - \nr
$  W t \bar t      $ & 0.24  & 0.02  & - \endtable
\hrule \vskip .04in \hrule
\endinsert

\TABLE\efftable{}
\topinsert
\titlestyle{\twelvepoint
Table~\efftable : $WW$ leptonic branching ratios, and the
efficiencies of jet-tagging and vetoing for the $\wl\wl$ signal
at the SSC (LHC).}
\bigskip
\thicksize=0pt
\hrule \vskip .04in \hrule
\begintable
$Z Z $    & branching ratio & tag only & veto plus tag \cr
$ {}              $ & 0.45\%& 59\% (49\%) & - \cr
$W^+ W^-$ & branching ratio & veto only & veto plus tag \cr
$  {}             $ & 4.7\% & 57\% (40\%) & 38\% (24\%) \cr
$W^+ Z$   & branching ratio & veto only & veto plus tag \cr
$ {}              $ & 1.5\% & 75\% (48\%) & 40\% (20\%) \cr
$W^+ W^+$ & branching ratio & veto only & veto plus tag \cr
$  {}             $ & 4.7\% & 69\% (58\%) & - \endtable

\hrule \vskip .04in \hrule
\endinsert

In Table~\cuttable\ we list the kinematic cuts used in our study
at the SSC (and LHC in parentheses).
In Tables~\smssctable a and~\smssctable b we present the cross sections
obtained in the SM from the electroweak processes at $\mh=1$ TeV and
at $\mh=0.1$ TeV, as well as those for the $q\anti q$ annihilation
continuum pair production (QCD) reactions.  The results in the
``leptonic cuts only'' column are those obtained by imposing
only the leptonic cuts of Table~\cuttable, including the
back-to-back cuts in the $\wp\wm$ and $\wpm\wpm$ channels.
In the next two columns, the cross sections
obtained after imposing jet-tagging and/or vetoing, {\it in addition
to the leptonic cuts}, are given.  The efficiencies at the SSC and LHC for
the signal are obtained
by taking the difference between the $\mh=1$ TeV and $\mh=0.1$ TeV results.
The branching ratios for each leptonic channel
and the efficiencies for the signal when performing jet-tagging
and/or vetoing (with leptonic cuts already imposed)
are summarized in Table~\efftable.

For other models of $\wl\wl$ interactions, we will proceed as follows.
We first compute the cross sections for $\wl\wl$ production in a given model
by using the Effective $W$ Approximation (EWA) \refmark{\effwtwo,\effwthree}
and the Equivalence Theorem (ET) \refmark{\lqt-\et}.
In using the EWA we compute total cross sections ignoring
all jet observables.   To assess the inaccuracies that
might arise as a result of these approximations, we will make a
detailed comparison between the EWA and ET computations
and the exact SM calculation in section 4.
To implement the lepton cuts, including back-to-back requirements
in the $\wp\wm$ and $\wpm\wpm$ channels, we decay
the final $\wl$'s according to the appropriate angular distributions.
The results will differ from the exact calculation to the extent that
lepton cut efficiencies depend upon the $p_T$ of the $WW$ system.
For the cuts employed, a comparison between the exact and EWA
lepton cut efficiencies is made in section~4 for the 1 TeV SM Higgs case,
and good agreement is found.
To obtain cross sections in the EWA approximation that include
the jet-tagging and jet-vetoing cuts,
we will simply multiply the cross sections calculated from EWA by the
net jet-tagging and/or jet-vetoing
efficiency for each channel as computed for the $\wl\wl$ signal
in the exact SM calculation with a 1 TeV Higgs boson.
We believe that this procedure should be fairly accurate. Indeed,
the kinematics of the jets in the
signal events are determined by the kinematics of
the initial $\wl$'s that participate in the $\wl\wl$ scattering
process.  These kinematics are independent of the
strong $\wl\wl$ scattering amplitude.

\chapter {Beyond the Standard Model}

\REF\jab{
For a review, see {\it e. g. }, J. Bagger, in R.~K.~Ellis, C.~T.~Hill, and
J.~D.~Lykken, eds., {\it Perspectives in the Standard Model},
World Scientific, 1992.}

In this section we present a variety of models that unitarize
the $W_LW_L$ scattering amplitude.  We start by reviewing the
Standard Model, and then discuss other possibilities
that are consistent with all the data to date \refmark{\jab}.

Let us begin by recalling that in the Standard Model, the $W_LW_L$
scattering amplitudes are unitarized by exchange of a spin-zero resonance,
the Higgs particle $H$.  The Higgs boson
is contained in a complex scalar doublet,
$$ \Phi\ =\ (v + H) \exp(2i w^a \tau^a/v)\ ,
\eqn\threeone
$$
where the $\tau^a$ are the generators of $SU(2)$, normalized so that
$\Tr \tau^a\tau^b = \delta^{ab} /2 $.  The four components of
$\Phi$ contain three would-be Goldstone bosons $w^a$
and the Higgs particle $H$.  In the Standard Model, the Higgs
potential,
$$
\V\ =\ {\lambda\over 16}\,
\bigg[ {\rm Tr}\,\big(\Phi^\dagger\Phi - v^2\big)\bigg]^2,
\eqn\threetwo
$$
is invariant under a rigid $SU(2)_L \times SU(2)_R$ symmetry,
$$
\Phi\ \rightarrow\ L\,\Phi\,R^\dagger\ ,
\eqn\threethree
$$
with $L,R \in SU(2)$.  The vacuum expectation value
$$
\langle\Phi\rangle \ =\  v\ ,
\eqn\threefour
$$
breaks the symmetry to the diagonal $SU(2)$.  In the perturbative limit,
it also gives mass to the Higgs boson,
$$
m_H\ =\ \sqrt{2 \lambda}\, v\ ,
\eqn\threefive
$$
where $v=246$ GeV.

In the Standard Model, the diagonal $SU(2)$ symmetry is broken only
by terms proportional to the hypercharge coupling $g'$ and the
up-down fermion mass splittings.  It is responsible for the
successful mass relation
$$
M_W\ =\ M_Z\ \cos\theta\ ,
\eqn\threesix
$$
where $\theta$ is the weak mixing angle; $M_W$ and $M_Z$ are the masses of
$W^\pm$ and $Z$, respectively.
The four components of $\Phi$ split into a triplet $w^a$ and a singlet
$H$ under the $SU(2)$ symmetry.  In analogy to the chiral symmetry of
QCD, we call the unbroken $SU(2)$ ``isospin.''

At high energies, the scattering of longitudinally polarized $W$ particles
can be approximated by the scattering of the would-be Goldstone bosons
$w^a$ \refmark{\lqt-\et}.
For the Standard Model, this is a calculational simplification, but for other
models it is a powerful conceptual aid as well.  For example, if one
thinks of the would-be Goldstone fields in analogy with the pions of
QCD, one expects the $W_LW_L$ scattering amplitudes to be unitarized
by a spin-one, isospin-one vector resonance, like the techni-rho.
Alternatively, if one thinks of the Goldstone fields in terms of the
linear sigma model, one expects the scattering amplitudes to be unitarized
by a spin-zero, isospin-zero scalar field like the Higgs boson.

In this paper, we are interested in the strongly interacting longitudinal
$W$'s in the TeV region.  We will ignore the gauge couplings
and the up-down fermion mass splittings.
Therefore, the SU(2) ``isospin'' is conserved.
The $W_LW_L$ scattering amplitudes can then be written in terms of isospin
amplitudes, exactly as in QCD.  If we assign isospin indices
as follows,
$$
W^a_L\ W^b_L\ \rightarrow\  W^c_L\ W^d_L\ ,
\eqn\threeseven
$$
then the scattering amplitude is given by
$$ \CM(W^a_LW^b_L\rightarrow W^c_LW^d_L)\ =\ A(s,t,u)\delta^{ab}
\delta^{cd}\ +\ A(t,s,u)\delta^{ac}\delta^{bd}\ +
\ A(u,t,s)\delta^{ad}\delta^{bc}\ ,
\eqn\threeeight
$$
where $a,b,c,d=1,2,3$. (We use $W_L$ to denote either $W_L^\pm$
or $Z_L$, where $W_L^\pm=(1/\sqrt{2})(W_L^1 \mp iW_L^2)$ and $Z_L=W_L^3$.)
All the physics of $W_LW_L$ scattering is contained in the
amplitude functions $A$.

Given the amplitude functions, the physical amplitudes for
boson-boson scattering are given as follows,
$$\eqalign{
\CM(W^+_LW^-_L\rightarrow Z_LZ_L)\ =&\ A(s,t,u) \cr
\CM(Z_LZ_L\rightarrow W^+_LW^-_L)\ =&\ A(s,t,u) \cr
\CM(W^+_LW^-_L\rightarrow W^+_LW^-_L)\ =&\ A(s,t,u)\ +\ A(t,s,u) \cr
\CM(Z_LZ_L\rightarrow Z_LZ_L)\ =&\ A(s,t,u)\ +\ A(t,s,u)\ +\ A(u,t,s) \cr
\CM(W^\pm_LZ_L\rightarrow W^\pm_LZ_L)\ =&\ A(t,s,u) \cr
\CM(W^\pm_LW^\pm_L\rightarrow W^\pm_LW^\pm_L)\ =&\ A(t,s,u)\ +
\ A(u,t,s)\ .\cr}
\eqn\threenine
$$
In these expressions, the amplitudes do not include the
symmetry factors for identical particles.

The isospin amplitudes $T(I)$, for isospin $I$, are given by
$$\eqalign{
T(0)\ & =\ 3\,A(s,t,u)\ +\ A(t,s,u)\ +\ A(u,t,s)  \cr
T(1)\ & =\ A(t,s,u)\ -\ A(u,t,s)  \cr
T(2)\ & =\ A(t,s,u)\ +\ A(u,t,s) \ .  \cr}
\eqn\threeten
$$
In terms of the isospin amplitudes, the physical scattering
amplitudes can be written
$$\eqalign{
\CM(W^+_LW^-_L\rightarrow Z_LZ_L)\ =&\  {1\over 3}[ T(0)\ -\ T(2)]  \cr
\CM(Z_LZ_L\rightarrow W^+_LW^-_L)\ =&\  {1\over 3}[ T(0)\ -\ T(2)]   \cr
\CM(W^+_LW^-_L\rightarrow W^+_LW^-_L)\ =&\  {1\over6}[2T(0)\ +\ 3T(1)\ +
\ T(2)] \cr
\CM(Z_LZ_L\rightarrow Z_LZ_L)\ =&\  {1\over3}[T(0)\ +\ 2T(2)]  \cr
\CM(W^\pm_LZ_L\rightarrow W^\pm_LZ_L)\ =&\  {1\over2}[T(1)\ +\ T(2)]  \crr
\CM(W^\pm_LW^\pm_L\rightarrow W^\pm_LW^\pm_L)\ =&\  T(2)\ .\cr}
\eqn\threemore
$$
Again, these amplitudes do not include the symmetry factors
for identical particles.

For the Standard Model, the amplitude functions are easy to
work out.  They can be expressed by
$$
\eqalign{
A(s,t,u)\ &=\ {-m_H^2\over v^2}\left(1 + {m_H^2\over s-m_H^2 +
i m_H \Gamma_H \theta (s)}\right)\ ,}
\eqn\threetwelve
$$
where $m_H$ and $\Gamma_H = 3 m^3_H/32 \pi v^2$ are the
mass and width of the Higgs boson; $\theta (s)$ is the step-function
which takes the value one for $s > 0$ and zero otherwise.
\REF\rwwll{C.--P. Yuan, Nucl. Phys. {\bf B310} (1988) 1.}
\REF\kane{G.L.~Kane and C.-P.~Yuan, Phys.~Rev. {\bf D40}, 2231 (1989).}
\REF\stuart{S. Willenbrock and G. Valencia,  Phys. Lett. {\bf B247}, 341
(1990); Phys. Rev. {\bf D46}, 2247 (1992).}
Note that we have included a Breit-Wigner width for the Higgs particle
in the $s$-channel.  This is a violation of the equivalence theorem,
which causes an increase in the rate in the resonant channels
\refmark{\rwwll-\stuart}.
More discussion of this violation will appear later.
We have not included the width in the non-resonant channels.

The Standard Model has the advantage that it is a renormalizable
theory, and that all amplitudes are perturbatively
unitary so long as $m_H$ is not too large.
Of course, at $\mh=1$ TeV, some small amount of unitarity violation
occurs near the resonance.  Nonetheless, the signal rates we
obtain for $\mh=1$ TeV provide a first characterization of what
one might expect in the case where $\wl\wl$ interactions become strong.

There are many other models that provide alternative
descriptions of electroweak symmetry breaking.
(We shall only consider models without any open inelastic channels
in $W_L W_L$ scattering.)  Many of these models
are effective theories, based on nonrenormalizable chiral lagrangians
for the $WW$ sector.  These models must be understood in the context
of an energy expansion. Generally, such an expansion does
not provide a unitary description for all energies. This is simply because
the effective lagrangian does not make explicit the new physics that must
appear at some scale $\Lambda$, well above the $WW$ mass region where it
is to be employed. For the purposes of this paper, we must ensure that
the effective theories are unitary for the $WW$ masses of interest.

To check unitarity, we first write the scattering amplitudes in
terms of the isospin amplitudes $T(I)$.  We then expand in partial waves
according to the usual formula
$$\eqalign{
T(I)\ &=\ 32\pi\sum_{\ell=0}^\infty (2\ell+1)P_\ell(\cos\theta)\ali\ ,  \cr
\ali\ &=\ {1\over 64\pi}\int^1_{-1}d(\cos\theta )P_\ell(\cos\theta )T(I)\ .
\cr}
\eqn\threethirteen
$$
Two-body elastic unitarity is equivalent to the statement
$| \ali - i/2 | = 1/2$.  In this paper we will require $\Re \ali
< 1/2$ as our unitarity condition.

Among possible alternative models, there are several distinctions we can
make.  The first is whether or not a particular model is resonant
in the $W_LW_L$ channel.  If it is resonant, the model can be classified
by the spin and isospin of the resonance.  If it is not, the analysis
is more subtle.  Nonetheless, we shall see that all possibilities can be
described in terms of two parameters.  In this work, we will restrict our
attention to models with spin-zero, isospin-zero
resonances (like the Higgs boson), and spin-one, isospin-one resonances
(like the techni-rho resonance), and nonresonant models.

\section{Spin-zero, Isospin-zero Resonances}

\subsection{3.1.1 The $O(2N)$ Model}

The first model we discuss represents an attempt to describe the
Standard Model Higgs in the nonperturbative domain.  In the
perturbatively-coupled Standard Model, the mass of the Higgs is
proportional to the square root of the scalar self-coupling
$\lambda$.  Heavy Higgs particles correspond to large values
of $\lambda$.  For $m_H \gsim$ 1 TeV, naive perturbation theory
breaks down, and one must take a more sophisticated approach.

One possibility for exploring the nonperturbative regime is to
exploit the isomorphism between $SU(2)_L \times SU(2)_R$ and
$O(4)$ \Ref\einhorn{M.~Einhorn, Nucl. Phys. {\bf B246}, 75 (1984).}.
Using a large-$N$ approximation, one can solve the $O(2N)$
model for all values of $\lambda$, to leading order in $1/N$.
The resulting scattering
amplitudes can be parameterized by the scale $\Lambda$ of the
Landau pole.  Large values of $\Lambda$ correspond to small couplings
$\lambda$ and relatively light Higgs particles.  In contrast, small
values of $\Lambda$ correspond to large $\lambda$ and describe the
nonperturbative regime.

The amplitude functions can be found via standard large-$N$
techniques.  In the limit $N \rightarrow \infty$, they
are \Ref\nachyuan{S.~Nachulich and C.--P.~Yuan, Phys. Lett. {\bf B293},
395 (1992).}
$$\eqalign{
%A(s,t,u)\ =&\ {16 \pi^2 s\over 16 \pi^2 v^2 -
%sN[ 2 + \ln (\Lambda^2/s)+i\pi]}, \cr
%A(t,s,u)\ =&\ {16 \pi^2 t\over 16 \pi^2 v^2 -
%tN[ 2 + \ln (-\Lambda^2/t)]}, \cr
%A(u,t,s)\ =&\ {16 \pi^2 u\over 16 \pi^2 v^2 -
%uN[ 2 + \ln (-\Lambda^2/u)]}, \cr}
A(s,t,u)\ =&\ {16 \pi^2 s\over 16 \pi^2 v^2 -
sN[ 2 + \ln (\Lambda^2/|s|)+i\pi \theta (s)]}\ ,}
\eqn\threefourteen
$$
where $\Lambda$ is the physical cutoff and $\theta (s)$ is the step-function
defined below Eq.~\threetwelve.  The scale of the
cutoff completely determines the theory.

It is not hard to show that the $W_LW_L$ scattering amplitudes
respect the unitarity condition for all energies $E \lsim \Lambda$.
In this paper we will take $N=2$ and
$\Lambda = 3$ TeV to characterize the
strongly-coupled Standard Model.
If we parameterize the position of the pole by its
``mass'' $m$ and ``width'' $\Gamma$ through the relation
$s=(m-{i \over 2} \Gamma)^2$, then $m\sim 0.8$ TeV and $\Gamma \sim 600$ GeV.

\subsection{3.1.2 The Chirally-Coupled Scalar Model}

The second model describes the low-energy regime of a technicolor-like
model whose lowest resonance is a techni-sigma.  The effective lagrangian
for such a resonance can be constructed using the techniques of Callan,
Coleman, Wess and Zumino
\Ref\ccwz{S.~Weinberg, Phys. Rev. {\bf 166}, 1568
(1968); S.~Coleman, J.~Wess and B.~Zumino, Phys. Rev. {\bf 177}, 2239 (1969);
C.~Callan, S.~Coleman, J.~Wess and B.~Zumino, Phys. Rev. {\bf 177},
2247 (1969); S.~Weinberg, Physica {\bf 96A}, 327 (1979).}.
The resulting lagrangian is consistent
with the chiral symmetry $SU(2)_L \times SU(2)_R$, spontaneously broken
to the diagonal $SU(2)$.

In this approach, the basic fields are $\Sigma = \exp (2 i w^a \tau^a/v)$
and a scalar $S$.  These fields transform as follows under $SU(2)_L \times
SU(2)_R$,
$$\eqalign{
\Sigma\ & \rightarrow \ L\,\Sigma\,R^\dagger\ , \cr
S\ & \rightarrow \ S\ .\cr}
\eqn\threefifteen
$$

This is all we need to construct the effective lagrangian.  To the
order of interest, it is given by
$$
\eqalign{
\L_{\rm Scalar}\ & =\ \ {v^2 \over 4}\,
\Tr \partial^\mu \Sigma^\dagger
\partial_\mu \Sigma \cr
& \ +\ {1\over2}\,\partial^\mu S \partial_\mu S \ -
\ {1\over2}\,M^2_S\,S^2 \cr
& \ +\ {1\over2}\,gv\,S\,\Tr \partial^\mu \Sigma^\dagger \partial_\mu
\Sigma\ +\ ... \ ,\cr}
\eqn\lscalar$$
where $M_S$ is the isoscalar mass, and $g$ is related to its partial
width into the Goldstone fields,
$$
\Gamma_S\ =\ {3 g^2 M^3_S\over 32 \pi v^2}\ .
\eqn\threeseventeen
$$

To this order, the lagrangian \lscalar\ is the most general
chirally-symmetric coupling of a spin-zero isoscalar resonance to the
fields $w^a$.  It contains two free parameters, which can be traded for
the mass and the width of the $S$.  For $g = 1$, the $S$ reduces
to an ordinary Higgs.  For $g \ne 1$, however, the $S$ is {\it not}
a typical Higgs.  It is simply an isoscalar resonance of arbitrary
mass and width.  In either case, one must be sure to check
that the scattering amplitudes are unitary up to the energy of
interest.

The tree-level scattering amplitude is easy to construct.  It has
two terms.  The first is a direct four-Goldstone coupling which
ensures that the scattering amplitude satisfies the
Low-Energy Theorems (LET)
\Ref\georgietal{S.~Weinberg, Phys. Rev. Lett. {\bf 17}, 616 (1966);
M.~S.~Chanowitz, M.~Golden, and H.~Georgi,
Phys. Rev. Lett. {\bf 57}, 2344 (1986);
Phys. Rev. {\bf D35}, 1490 (1987).}.
The second contains the contributions from the
isoscalar resonance.  Taken together, they give the full scattering
amplitude,
$$
\eqalign{
A(s,t,u)\ &=\ {s\over v^2}\ -\ \bigg({g^2 s^2 \over v^2}\bigg)
{1\over s - M_S^2 + i M_S \Gamma_S \theta (s)}\ . }
%A(t,s,u)\ &=\ {t\over v^2}\ -\ \bigg({g^2 t^2 \over v^2}\bigg)
%{1\over t - M_S^2}\ ,\crr
%A(u,t,s)\ &=\ {u\over v^2}\ -\ \bigg({g^2 u^2 \over v^2}\bigg)
%{1\over u - M_S^2}\ .\cr}
\eqn\threeeighteen$$
In what follows, we will choose $M_S = 1.0$ TeV, $\Gamma_S = 350$
GeV.  These values give unitary scattering amplitudes
up to 2 TeV.
(We use the Breit--Wigner prescription to handle the
$s$--channel resonance. Our criteria is to have all the partial
waves respect the unitarity condition up to 2 TeV except near the resonance;
the slight unitarity violation near the resonance is
due to the perturbative expansion of the width \refmark{\rwwll-\stuart}.)

\section{Spin-one, Isospin-one Resonances}

\subsection{3.2.1 The Chirally Coupled Vector Model}

\REF\BESS{R.~Casalbuoni, \it et al., \rm Phys. Lett.
{\bf B155}, 59 (1985); Nucl. Phys. {\bf B282}, 235 (1987);
see also M.~Bando, T.~Kugo, and K.~Yamawaki, Phys. Rep. {\bf
164}, 217 (1988), and references therein.}
\REF\BESSssc{R.~Casalbuoni, \it et al., \rm Phys. Lett.
{\bf B249}, 130 (1990); {\bf B253}, 275 (1991); J.~Bagger,
T.~Han and R.~Rosenfeld, in E.~Berger, ed., \it Research
Directions for the Decade, Snowmass 1990, \rm World
Scientific, 1992.}
\REF\besslimit{R.~Casalbuoni, \it et al., \rm Phys. Lett.
{\bf B269}, 361 (1991).}

This example provides a relatively model-independent description of
the techni-rho resonance that arises in most technicolor theories.
As above, one can use the techniques of nonlinear realizations to
construct the most general coupling consistent with chiral
symmetry \refmark{\ccwz,\BESS-\besslimit}.

To find the techni-rho lagrangian, we first parameterize the Goldstone
fields $w^a$ in a slightly different way,
$$
\xi\ =\ \exp(i w^a \tau^a/v)\ ,
\eqn\xidef
$$
so $\Sigma = \xi^2$.  We then represent an $SU(2)_L \times SU(2)_R$
transformation on the field $\xi$ as follows:
$$
\xi\ \rightarrow\ \xi^\prime\ \equiv
\ L\, \xi\, U^\dagger\ =\ U\, \xi\, R^\dagger\ .
\eqn\threetwenty
$$
Here $L,\ R$ and $U$ are $SU(2)$ group elements, and $U$ is a
(nonlinear) function of $L,\ R$ and $w^a$, chosen to restore
$\xi^\prime$ to the form \xidef.  Note that when $L = R$, $U = L
= R$ and the transformation linearizes.  This simply says that the
$w^a$ transform as a triplet under the diagonal $SU(2)$.

Given these transformations, one can construct the following currents,
$$\eqalign{
J_{\mu L}\ &=\ \xi^\dagger \partial_\mu \xi \ \rightarrow\ U J_{\mu L}
U^\dagger  \ +\ U \partial_\mu U^\dagger\ , \cr
J_{\mu R}\ &=\ \xi \partial_\mu \xi^\dagger \ \rightarrow\ U J_{\mu R}
U^\dagger  \ +\ U \partial_\mu U^\dagger\ .  \cr}
\eqn\jtrans
$$
The currents $J_{\mu L}$ and $J_{\mu R}$ transform as gauge fields under
transformations in the diagonal $SU(2)$.  As above, the
transformations linearize when $L=R=U$.

The transformations \jtrans\ inspire us to choose the techni-rho
transformation as follows,
$$
V_\mu \ \rightarrow\ U V_\mu U^\dagger
 \ +\ ig^{-1}\, U \partial_\mu U^\dagger\ .
\eqn\trhotrans
$$
In this expression, $V_\mu = V_\mu^a \tau^a$, and $g$ is the techni-rho
coupling constant.  When $L=R=U$, Eq.~\trhotrans\ implies
that the techni-rho transforms as an isotriplet of weak isospin.

Using these transformations, it is easy to construct the most general
lagrangian consistent with chiral symmetry.  We first write down the
currents
$$\eqalign{
\A_\mu\ &=\  J_{\mu L} \ -\ J_{\mu R}\ , \cr
\V_\mu\ &=\  J_{\mu L} \ +\ J_{\mu R}\ +\ 2ig V_\mu \ , \cr}
\eqn\threetwentyfive
$$
which transform as follows under an arbitrary chiral transformation,
$$\eqalign{
\A_\mu\ &\rightarrow\ U \A_\mu U^\dagger\ , \cr
\V_\mu\ &\rightarrow\ U \V_\mu U^\dagger \ .  \cr}
\eqn\threetwentysix
$$
Under parity (which exchanges $J_L$ with $J_R$ and leaves $V$
invariant), $\V$ is invariant, while $\A$ changes sign.  If we make
the additional assumption that the underlying dynamics conserve
parity, we are led to the following lagrangian,
$$
\eqalign{
\L_{{\rm Vector}}\ & =
%\ -\ {v^2 \over 4}\, \Tr \partial_\mu \Sigma^\dagger\partial^\mu \Sigma
\ -\ {1\over 4}\, V^a_{\mu\nu} V^{a\mu\nu}
\ -\ {1\over4}\,v^2\,\Tr \A_\mu \A^\mu
 \ -\ {1\over4}\,a v^2\,\Tr \V_\mu \V^\mu\ +\ ... \ ,\cr}
\eqn\threetwentyseven
$$
where $V^a_{\mu\nu}$ is the (nonabelian) field-strength for the
vector field $V^a_\mu$.  The dots in this equation denote terms with
more derivatives.  Up to a possible field redefinition, this is the
most general coupling of a techni-rho resonance to the Goldstone
bosons, consistent with $SU(2)_L \times SU(2)_R$ symmetry.

In this lagrangian, the parameter $v$ is fixed as before.  The
parameters $g$ and $a$, however, are free.  One combination is
determined by the mass of the techni-rho,
$$
M^2_V\ =\ a g^2 v^2\ ,
\eqn\threetwentyeight
$$
and another by its width into techni-pions (\ie\ Goldstone bosons),
$$
\Gamma_V \ =\ {a M_V^3 \over 192 \pi v^2} \ .
\eqn\threetwentynine
$$
Because of the chiral symmetry, these two parameters completely
define the theory.

As above, the scattering amplitude is easy to compute.
It contains a direct four-Goldstone-boson coupling, as well as the
isovector resonance.  One finds
$$\eqalign{
A(s,t,u)\ &=\ {s\over 4 v^2}\,\bigg(4 - 3\,a\bigg)
\ +\ { a M^2_V \over 4 v^2}\,\bigg[ {u-s \over t - M^2_V + i M_V \Gamma_V
\theta (t)}\crr
& \qquad\qquad  +\ {t-s \over u - M^2_V + i M_V \Gamma_V \theta (u)}
\bigg]\ . }
%A(t,s,u)\ &=\ {t\over 4 v^2}\,\bigg(4 - 3\,a\bigg)
%\ +\ { a M^2_V \over 4 v^2}\,\bigg[ {u-t \over s - M^2_V + i M_V \Gamma_V}
%\ +\ {s-t \over
%u - M^2_V} \bigg]\ ,   \crr
%A(u,t,s)\ &=\ {u\over 4 v^2}\,\bigg(4 - 3\,a\bigg)
%\ +\ { a M^2_V \over 4 v^2}\,\bigg[ {s-u \over t - M^2_V}\ +\ {t-u \over
%s - M^2_V + i M_V \Gamma_V} \bigg]\ .  \cr}
\eqn\vecamp
$$

In what follows we will choose
$M_\rho = 2.0$ TeV, $\Gamma_\rho = 700$ GeV and
$M_\rho = 2.5$ TeV, $\Gamma_\rho = 1300$ GeV.
These values preserve unitarity up to 3 TeV,
except for a small unitarity violation near the $s$-channel
resonance in the $a^1_1$ partial wave.  Additional constraints
can be found from precision measurements of the electroweak
parameters.  Our choices are consistent with current limits
\refmark{\besslimit}.

\section{Nonresonant models}

\REF\nrm{A.~Dobado, M.~Herrero and J.~Terron,
Z. Phys. {\bf C50}, 205 (1991); 465 (1991);
S.~Dawson and G.~Valencia, Nucl. Phys. {\bf B348}, 23 (1991);
Nucl. Phys. {\bf B352}, 27 (1991);
J.~Bagger, S.~Dawson and G.~Valencia, in E.~Berger, ed.,
\it Research Directions for the Decade, Snowmass 1990,
\rm World Scientific, 1992.}

\REF\cpt{J.~Gasser and H.~Leutwyler,
Ann. Phys. {\bf 158}, 142 (1984); Nucl. Phys.
{\bf B250}, 465 (1985); J.~Bagger, S.~Dawson and
G.~Valencia, Fermilab preprint PUB-92-75-T (1992),
and references therein.}

\REF\lukef{See, however, A.F.~Falk, M.~Luke and E.H.~Simmons,
Nucl. Phys. {\bf B365} (1991) 523.}

Effective field theories provide a useful formalism
for describing resonances in $W_LW_L$ scattering beyond the
Standard Model.  They also can be used to
describe nonresonant models in which the $W_LW_L$
scattering occurs below the threshold for resonance production.
The effective lagrangian description allows one to construct scattering
amplitudes that are consistent with crossing, unitarity and chiral
symmetry \refmark{\nrm}.

The most important effects at SSC energies can be found by considering
the lagrangian for the Goldstone fields,
$$\eqalign{
\L_{\rm Goldstone} \ & =  {v^2 \over 4}\,
\Tr \partial_\mu \Sigma^\dagger
\partial^\mu \Sigma \cr
 &\ +\ L_1\,\bigg({v \over \Lambda}\bigg)^2\,
 \Tr(\partial_\mu\Sigma^\dagger \partial^\mu \Sigma)
 \ \Tr(\partial_\nu\Sigma^\dagger \partial^\nu \Sigma) \cr
& \ +\ L_2\,\bigg({v \over \Lambda}\bigg)^2\,\Tr(\partial_\mu\Sigma^\dagger
\partial_\nu \Sigma)\ \Tr(\partial^\mu\Sigma^\dagger \partial^\nu \Sigma)\ ,
\cr}
\eqn\goldlag
$$
where $\Lambda \lsim 4 \pi v$ denotes the scale of the new physics.
To this order, this is the most general $SU(2)_L \times SU(2)_R$
invariant lagrangian for the Goldstone fields \refmark{\cpt,\lukef}.

The lagrangian \goldlag\ describes new physics at energies below the
mass of lightest new particles.  All the effects of the new physics
are contained in the coefficients of the higher-dimensional operators
built from the Goldstone fields.  To order $p^2$ in the energy expansion,
only one operator contributes, and its coefficient is universal.  To order
$p^4$, however, there are two additional operators that contribute to
$W_LW_L$ scattering.

To order $p^4$, the scattering amplitudes are given by
$$\eqalign{
A(s,t,u) \ &=  {s \over v^2}\ +
\ {1 \over 4 \pi^2 v^4}\ \biggl( 2\, L_1(\mu)\,s^2
\ +\ L_2(\mu)\, (t^2+u^2)\ \biggr)  \cr
&\ +\ {1 \over 16 \pi^2 v^4}\ \biggl[-{t \over 6}\,(s+2t){\rm log}
\biggl(-{t \over \mu^2}\biggr)\ -\ {u \over 6}\,(s+2u){\rm log}
\biggl(-{u \over \mu^2}\biggr)\  \cr
&\ -\ {s^2 \over 2}\, {\rm log}\biggl(-
{s \over \mu^2} \biggr)\biggr]\ ,}
%A(t,s,u) \ &=  {t \over v^2}\ +
%\ {1 \over 4 \pi^2 v^4}\ \biggl( 2\, L_1(\mu)\,t^2
%\ +\ L_2(\mu)\, (s^2+u^2)\ \biggr)  \cr
%&\ +\ {1 \over 16 \pi^2 v^4}\ \biggl[-{s \over 6}\,(t+2s){\rm log}
%\biggl(-{s \over \mu^2}\biggr)\ -\ {u \over 6}\,(t+2u){\rm log}
%\biggl(-{u \over \mu^2}\biggr)\  \cr
%&\ -\ {t^2 \over 2}\, {\rm log}\biggl(-
%{t \over \mu^2} \biggr)\biggr]\ \crr
%A(u,t,s) \ &=  {u \over v^2}\ +
%\ {1 \over 4 \pi^2 v^4}\ \biggl( 2\, L_1(\mu)\,u^2
%\ +\ L_2(\mu)\, (t^2+s^2)\ \biggr)  \cr
%&\ +\ {1 \over 16 \pi^2 v^4}\ \biggl[-{t \over 6}\,(u+2t){\rm log}
%\biggl(-{t \over \mu^2}\biggr)\ -\ {s \over 6}\,(u+2s){\rm log}
%\biggl(-{s \over \mu^2}\biggr)\  \cr
%&\ -\ {u^2 \over 2}\, {\rm log}\biggl(-
%{u \over \mu^2} \biggr)\biggr]\ ,\cr}
\eqn\goldamps
$$
where we have taken $\Lambda = 4 \pi v \sim 3.1$ TeV and
the $L_i(\mu)$ are the renormalized coefficients in the
effective lagrangian.
(${\rm log}(-s) = {\rm log}(s)-i \pi$, for $s > 0$.)
To this order, there are two types
of contributions.  The first is a direct coupling that follows
from the tree-level lagrangian.  The second is a one-loop correction
that must be included at order $p^4$.  The loop contribution
renormalizes the parameters $L_1$ and $L_2$, and gives finite
logarithmic corrections that cannot be absorbed into a redefinition
of the couplings.

The difficulty with this approach is that at SSC energies, the scattering
amplitudes violate unitarity between 1 and 2 TeV.  This indicates
that new physics is near, but there is no guarantee that new resonances
lie within the reach of the SSC. We choose to treat the uncertainties of
unitarization in three ways:

1)  We take $L_1(\mu) = L_2(\mu) = 0$ and ignore the loop-induced
logarithmic corrections to the scattering amplitudes.  The resulting
amplitudes are universal in the sense that they depend only on $v$.
They reproduce the low-energy theorems of pion dynamics.
We unitarize these amplitudes by cutting off the partial waves when
they saturate the bound $|\ali| < 1$.   This is the original model
considered by Chanowitz and Gaillard \refmark{\changail},
so we call it {\it LET CG}.

2)  For comparison, we consider another model in which we take $L_1(\mu)
= L_2(\mu) = 0$ and ignore the loop-induced logarithmic corrections.
This time, however, we unitarize the scattering amplitudes using a
``$K$-matrix;''  that is, we replace the partial wave amplitudes $\ali$
by $\tli$, where
$$
\tli\ =\ {\ali \over 1-i\ali }\ .
\eqn\tamps
$$
We call this model {\it LET K}.

3)  The third nonresonant model we consider includes the full
${\cal O}(p^4)$ amplitude presented above.  By varying the parameters
$L_1(\mu)$ and $L_2(\mu)$, one can sweep over all possible nonresonant
physics.  In particular, one can search for a region where unitarity
violation is delayed up to 2 TeV.  Scanning the ($L_1(\mu), L_2(\mu)$)
parameter space, one finds that the values
$$\eqalign{
L_1(\mu)\ &=\ -0.26 \cr
L_2(\mu)\ &=\ +0.23 \ ,\cr}
\eqn\loneltwo
$$
measured at the renormalization scale $\mu=1.5$ TeV, delay unitarity
breakdown until 2 TeV.  With these parameters, the amplitudes
\goldamps\ are unitary, chiral and crossing-symmetric for energies up
to 2 TeV.  Beyond 2 TeV, the partial waves are no longer unitary.
In order to compare with the total event rates in the
other models, we unitarize the scattering
amplitudes using the $K$-matrix prescription,
so we call this model {\it DELAY K}.
Note that only the real part of $a^I_\ell$,
from \threethirteen\ and \goldamps,
is used to obtain the unitarized partial wave amplitude $t^I_\ell$.

In what follows we use these models to represent new physics
that is not resonant at SSC and LHC energies.

\chapter{ Numerical Results}

\REF\mftg{J. G. Morfin and W.-K. Tung, Z. Phys. {\bf C52}, 13 (1991). }
\REF\changuess{M. Berger and M. Chanowitz, Phys.\ Lett.\ {\bf B267}, 416
(1991).}
\REF\hvw{T. Han, G. Valencia, and S. Willenbrock, Phys.\ Rev.\ Lett.\
{\bf 69}, 3274 (1992).}
\REF\ewat{J.~F.~Gunion, J.~Kalinowski, A.~Tofighi-Niaki, A.~Abbasabadi,
and W.~Repko, in {\it Physics of the Superconducting Supercollider},
Snowmass, Colorado (1986); Gunion \etal, Ref. \effwthree;
Abbasabadi \etal, Ref. \effwthree.}

We now turn to the background and signal results for the models discussed in
sections 2 and 3. We begin by briefly summarizing our procedures
and assumptions and then estimate the overall systematic error
associated with the event rates to be presented.

In all our results, we use the
leading-order parton distributions of Morfin and Tung \refmark{\mftg},
and include only the first four quark flavors as partons.
In particular, we ignore the bottom quark as
an initial-state parton.  In computing signal
event rates, we evaluate the parton distribution functions
at the scale $\mw$. As discussed in Ref.~{\changuess},
the agreement between the exact $W_LW_L$ production rates and those predicted
by the effective--$W$ approximation is best for this (natural) choice of scale.
In a recent analysis that includes the next-order QCD
corrections \refmark{\hvw}, it was found that for this choice of scale
the QCD corrections to
the $WW$ scattering processes are very small, which indicates that the
current tree-level calculations for the signals are rather reliable.
The scales used in the background calculations depend upon the process,
and are given in the references quoted in section 2.
In general, the
choices of scale are either strongly motivated on theoretical grounds
or are those leading to the smallest higher--order corrections. To estimate
the systematic error that is associated with our background rates
because of scale choice and higher--order corrections,
we need higher-order calculations which
are not available at present.  In the TeV region, however, the typical scales
are large.  This makes the strong coupling small and
takes the parton distribution functions
into regimes of relatively large momentum fraction, which are well-represented
by experimental data.

\TABLE\comptable{}
\topinsert
\titlestyle{\twelvepoint
Table~\comptable : SM signal cross section comparison between
the Subtraction results of Eq.~\twothirteen\
and the EWA/ET results ($m_H=$1 TeV) at the
SSC, $\sqrt{s} = 40$ TeV, in units of fb. Only the leptonic cuts in Table~1
are imposed.}
\bigskip
\thicksize=0pt
\hrule \vskip .04in \hrule
\begintable
$Z Z$ & Subtraction  & EWA/ET SM  \cr
$M_{ZZ} > 0.5$ & 1.0 & 1.9   \nr
$M_{ZZ} > 1.0$ & 0.52 & 0.69 \nr
$M_{ZZ} > 1.5$ & 0.06 & 0.08 \cr
$W^+ W^-$ & Subtraction  & EWA/ET SM  \cr
$M_{\ell\ell} > 0.25$ & 9.0 & 12.6 \nr
$M_{\ell\ell} > 0.5$ & 7.8 & 12.1 \nr
$M_{\ell\ell} > 1.0$ & 0.68 & 1.0 \cr
$W^+ Z$ & Subtraction  & EWA/ET SM  \cr
$M_T > 0.5$ & 0.29 & 0.33 \nr
$M_T > 1.0$ & 0.17 & 0.13 \nr
$M_T > 1.5$ & 0.08 & 0.05 \cr
$W^+ W^+$ & Subtraction  & EWA/ET SM  \cr
$M_{\ell\ell} > 0.25$ & 0.90 & 0.93  \nr
$M_{\ell\ell} > 0.5$ & 0.47 & 0.46  \nr
$M_{\ell\ell} > 1.0$ & 0.10 & 0.10  \endtable
\hrule \vskip .04in \hrule
\endinsert

To check the accuracy of employing the effective--$W$ approximation in
combination with the equivalence theorem  (EWA/ET),
we present a comparison in Table~\comptable.
For the test case of the Standard Model with a 1 TeV Higgs boson,
the agreement between this approximate technique (EWA/ET) and the
``subtraction'' result using the full SM matrix element calculation
(Subtraction) is reasonably good and generally
becomes best at large invariant mass of the final state $WW$ pair.
However, in the $\wp\wp$ final state the agreement is excellent for
all mass cuts examined. For other channels,
the discrepancy at lower $\mww$ potentially derives from two sources:
(a) use of the  EWA for the longitudinally polarized $W$--boson scattering
amplitudes, and (b) unavoidable inconsistencies associated with implementing
the equivalence theorem, leading to a difference between EWA/ET and EWA/LL.
(The EWA/LL approximation is that in which the EWA is employed
in conjunction with the full longitudinal $W$--boson scattering amplitudes.)
Regarding (a), we note that
the derivation of the EWA intrinsically relies on $\mww$ being large. Thus,
it is natural that some deviation between the Subtraction and EWA
computations for the $W_LW_L$ final state could appear at low $WW$
invariant mass.  However, the close agreement between the exact
and EWA/ET results for the $\wp\wp$ channel suggest that this source
of deviation is quite small.  This is because $\wp\wp$, being non-resonant,
does not suffer from difficulties of type (b).
Indeed, good agreement between exact and EWA/LL calculations for the (opposite
charge) $\wp\wm$ channel has been found in earlier work \refmark{\ewat}
where the ET approximation was not employed.
The main difference of type (b), \ie\ between EWA/LL and EWA/ET,
arises from our procedure of
employing the Breit--Wigner prescription for $s$-channel resonances.
As discussed below, if this procedure is employed (in the SM)
for both a direct calculation of $W_LW_L\rta W_LW_L$ using true $W$--boson
fields and in
the ET calculation of the same process, a large discrepancy
is found for $\mww$ below the Higgs mass.
The net deviation between the full matrix element calculation,
as defined in Eq.~\twothirteen, and the EWA/ET is displayed in
Table \comptable.

Let us discuss briefly the inconsistencies associated with
employing the Breit--Wigner
prescription for the $s$-channel resonances in the scattering amplitudes.
These were studied in the case of a heavy SM Higgs boson in Ref.~\kane.
As already noted, the Breit--Wigner procedure for
putting the width into the directly computed
$W_LW_L\rta W_LW_L$ amplitudes with true longitudinally
polarized gauge bosons does not yield the same result
as the identical procedure in the equivalence-theorem calculation.
However, it is easily demonstrated that this violation of the equivalence
theorem is higher--order in the perturbative expansion.
When the width is small, the perturbative expansion is valid and
the violation is tiny.
For large width, the Breit--Wigner procedure yields a significant
violation of the equivalence theorem for $\mww$ below the resonance
mass, but an unambiguous treatment is impossible
because the perturbative expansion is breaking down.
Our procedure can simply be viewed as defining a particular model
for $W_LW_L$ production.
The second effect of adding the width through the Breit--Wigner prescription
is to give a small violation of unitarity in the partial waves which contain
the resonance. This can again be traced to a breakdown of the
perturbative calculation and the Breit--Wigner induced violation of
the equivalence theorem. This small violation of unitarity near the resonance
can be safely ignored. Much more important is our
demand that the unitarity conditions hold away
from the resonance, in particular up to the highest $\mww$ scale of interest.
Indeed, our cuts automatically emphasize the large $\mww$ region
in which the Breit--Wigner procedure becomes immaterial.
Thus, as already stated, we see no reason to anticipate large errors
in our signal rates in the large $M(WW)$ region.

Finally, we remind the reader that our {\it signal} is defined as
the number of $W_LW_L$ pairs produced in any given channel.
In general, this is not the same as one would obtain by plotting
events as a function of $\mww$ and then subtracting a smooth background
under some ``bump'' in the distribution. Even in the fully
reconstructable four-lepton final state of the $ZZ$
channel, a spin--zero isospin--zero resonance is significantly
above the continuum background at high $M(ZZ)$, well beyond the obvious
bump in the distribution. Indeed, in non-resonant channels, such as $\wp\wp$,
there is no visible bump in $\mww$. And,
for most of the final states considered, the missing neutrinos make full
reconstruction of the $WW$ mass impossible in any case.
The ability to detect the signal will thus ultimately depend upon
the accuracy with which the expected rate for $WW$ production in
the $LT$ and $TT$ polarization modes can be computed.  As we have already
emphasized, the SM computation (with a light Higgs boson)
gives an accurate result for this rate.

The signal and background rates that we shall quote are only
as good as the parton distributions and parton-level Monte Carlo
programs employed.  Significant improvements in both will be made
once data from HERA is available. Also, the
required parton distributions will be determined
to high accuracy by other high-$p_T$ and Drell-Yan pair
measurements at the SSC or LHC, and implementation of the cuts
by each detector collaboration will become very
well understood as experience with the apparatus accumulates and
the full hadronic-level Monte Carlo simulations are fine-tuned at SSC/LHC
energies.  In particular, the background levels in the various channels
should become sufficiently well-determined that any significant
$LL$ ({\it i.e.} $W_LW_L$) excess
will be evident. Although at present we are confined to
parton-level predictions, our results should give a reliable
indication of the ultimate rates for the background and
signal that can be achieved in each model after appropriate cuts.

\TABLE\ssclhctable{}
\topinsert
\titlestyle{\twelvepoint
Table \ssclhctable a: Event rates per SSC-year, assuming $\mt=140\gev$,
$\sqrt s=40\tev$, and an annual luminosity of $10 \fbi$. Cuts are listed
in Table~1.}
\bigskip
\thicksize=0pt
\hrule \vskip .04in \hrule
\begintable

$Z Z$ \hfill | Bkgd. & SM & Scalar & $O(2N)$ & Vec 2.0 & Vec 2.5 & LET CG &
   LET K & Delay K  \cr

$M_{ZZ} > 0.5$ \hfill | 1.0 & 11 & 6.2 & 5.2 & 1.1 & 1.5 & 2.6 & 2.2 & 1.6\nr
$M_{ZZ} > 1.0$ \hfill | 0.3 & 4.1 & 2.6 & 2.0 & 0.4 & 0.7 & 1.6 & 1.3 & 0.8\nr
$M_{ZZ} > 1.5$ \hfill | 0.1 & 0.5 & 0.2 & 0.5 & 0.1 & 0.3 & 0.9 & 0.6 & 0.4\cr

$W^+ W^-$ \hfill | Bkgd. & SM & Scalar & $O(2N)$ & Vec 2.0 & Vec 2.5 & LET
CG &
   LET K & Delay K  \cr

$M_{\ell\ell} > 0.25$ \hfill | 21 & 48 & 30 & 24 & 15 & 12 & 16 & 12 &
11  \nr
$M_{\ell\ell} > 0.5$ \hfill |  17 & 46 & 29 & 23 & 15 & 12 & 15 & 11
& 11  \nr
$M_{\ell\ell} > 1.0$ \hfill |  3.6& 3.8 & 1.1 & 2.7 & 6.5 & 4.9 & 5.3 &
3.6 & 4.6  \cr

$W^+ Z$ \hfill | Bkgd. & SM & Scalar & $O(2N)$ & Vec 2.0 & Vec 2.5 & LET CG
&  LET K & Delay K  \cr

$M_T > 0.5$ \hfill | 2.5 & 1.3 & 1.8 & 1.5 & 9.5 & 6.2 & 5.8 & 4.9 & 6.0  \nr
$M_T > 1.0$ \hfill | 0.8 & 0.5 & 0.8 & 0.7 & 7.9 & 4.7 & 4.1 & 3.3 & 4.6  \nr
$M_T > 1.5$ \hfill | 0.3 & 0.2 & 0.2 & 0.3 & 5.5 & 3.2 & 2.6 & 1.9 & 3.2  \cr

$W^+ W^+$ \hfill | Bkgd. & SM & Scalar & $O(2N)$ & Vec 2.0 & Vec 2.5 & LET
CG &  LET K & Delay K  \cr

$M_{\ell\ell} > 0.25$ \hfill | 3.5 & 6.4 & 8.2 & 7.1 & 7.8 & 11 & 25 &
21 & 15  \nr
$M_{\ell\ell} > 0.5$ \hfill | 1.5 & 3.2 & 4.2 & 3.9 & 3.8 & 6.3 & 19 &
15 & 11  \nr
$M_{\ell\ell} > 1.0$ \hfill | 0.2 & 0.7 & 0.6 & 0.9 & 0.5 & 1.2 & 7.6 &
5.2 & 5.2  \endtable

\hrule \vskip .04in \hrule

\endinsert

\topinsert
\titlestyle{\twelvepoint
Table \ssclhctable b: Event rates per LHC-year, assuming $\mt=140\gev$,
$\sqrt s=16\tev$, and an annual luminosity of $100 \fbi$. Cuts are listed
in Table~1.}
\bigskip
\thicksize=0pt
\hrule \vskip .04in \hrule
\begintable

$Z Z$ \hfill | Bkgd. & SM & Scalar & $O(2N)$ & Vec 2.0 & Vec 2.5 & LET CG &
   LET K & Delay K  \cr

$M_{ZZ} > 0.5$ \hfill | 1.0 & 14 & 7.5 & 6.4 & 1.4 & 1.7 & 2.5 & 2.2 & 1.8 \nr
$M_{ZZ} > 1.0$ \hfill | 0.1 & 3.9 & 2.7 & 1.8 & 0.4 & 0.6 & 1.1 & 0.9 & 0.6 \nr
$M_{ZZ} > 1.5$ \hfill | 0.0 & 0.3 & 0.1 & 0.3 & 0.1 & 0.2 & 0.4 & 0.3 & 0.2 \cr

$W^+ W^-$ \hfill | Bkgd. & SM & Scalar & $O(2N)$ & Vec 2.0 & Vec 2.5 & LET
CG &    LET K & Delay K  \cr

$M_{\ell\ell} > 0.25$ \hfill | 18 & 40 & 26 & 19 & 8.0 & 6.8 & 9.2 & 7.2 &
6.2  \nr
$M_{\ell\ell} > 0.5$ \hfill |  15 & 32 & 21 & 16 & 7.4 & 6.1 & 8.3 & 6.3
& 5.5  \nr
$M_{\ell\ell} > 1.0$ \hfill | 2.5 & 1.3 & 0.4 & 1.0 & 2.4 & 1.6 & 1.7 &
1.1 & 1.2  \cr

$W^+ Z$ \hfill | Bkgd. & SM & Scalar & $O(2N)$ & Vec 2.0 & Vec 2.5 & LET CG
&  LET K & Delay K  \cr

$M_T > 0.5$ \hfill | 2.4 & 1.0 & 1.4 & 1.1 & 4.8 & 3.2 & 3.2 & 2.9 & 3.0  \nr
$M_T > 1.0$ \hfill | 0.3 & 0.3 & 0.4 & 0.3 & 3.3 & 1.8 & 1.6 & 1.4 & 1.7  \nr
$M_T > 1.5$ \hfill | 0.1 & 0.1 & 0.1 & 0.1 & 2.0 & 1.0 & 0.8 & 0.6 & 0.9 \cr

$W^+ W^+$ \hfill | Bkgd. & SM & Scalar & $O(2N)$ & Vec 2.0 & Vec 2.5 & LET
CG &   LET K & Delay K  \cr

$M_{\ell\ell} > 0.25$ \hfill | 6.2 & 9.6 & 12 & 10 & 12 & 16 & 27 & 24 & 16
\nr
$M_{\ell\ell} > 0.5$ \hfill | 1.7 & 3.7 & 5.2 & 4.3 & 4.8 & 7.3 & 16 &
14 & 8.3  \nr
$M_{\ell\ell} > 1.0$ \hfill | 0.2 & 0.4 & 0.5 & 0.6 & 0.4 & 1.0 & 4.2 &
2.9 & 2.3  \endtable

\hrule \vskip .04in \hrule
\endinsert

\FIG\fthree{  Invariant mass distributions for the
``gold-plated'' leptonic final states that arise
from the processes $pp \rightarrow ZZX$,
$pp \rightarrow W^+W^-X$, $pp \rightarrow W^+ZX$ and
$pp \rightarrow W^+W^+X$, for $\sqrt s = 40$ TeV
and an annual SSC luminosity of 10~fb$^{-1}$.
The longitudinally-polarized signal is plotted above the
summed background. The mass variable of
$x$-axis is in units of GeV, and the bin size is 50 GeV.
\nextline
a)  SM with a 1 TeV Higgs boson;
\nextline
b) $O(4)$ model with $\Lambda = 3$ TeV;
\nextline
c) Chirally coupled scalar with $M_S = 1$ TeV, $\Gamma_S = 350$ GeV;
\nextline
d) Chirally coupled vector with $M_V = 2$ TeV, $\Gamma_V = 700$ GeV;
\nextline
e) Chirally coupled vector with $M_V = 2.5$ TeV, $\Gamma_V = 1300$ GeV;
\nextline
f) Nonresonant model unitarized following Chanowitz and Gaillard;
\nextline
g) Nonresonant model unitarized by the $K$-matrix prescription;
\nextline
h) ${\cal O}(p^4)$ nonresonant model with delayed unitarity
violation, unitarized by the
{\hbox {$K$-matrix}} prescription.}
%
%Each plot
%contains three histograms, overlaid one on top of the
%other.  The white histogram always contains the background,
%that is, the SM with a 100 GeV Higgs.  In (a) and (b) the
%light histogram includes the signal plus background for the
%$O(2N)$ model, while the dark histogram contains the signal
%plus background for the SM with a 1 TeV Higgs.  In (c) the
%light histogram contains the contribution from a vector
%isovector resonance with $M = 2$ TeV and $\Gamma = 700$ GeV.
%The dark histogram represents the result for $M = 2.5$ TeV
%and $\Gamma = 1300$ GeV.  Figure (d) includes the signal plus
%background from two nonresonant models.  The light histogram
%contains the LET CG; the dark, DELAY K. }

We summarize our results in Tables \ssclhctable a and
\ssclhctable b for the SSC and the LHC,
respectively. These tables give the event rates for the summed
background and for the signal in each of the models as a function of the
mass cut placed on the final state.  The particular type of mass cut
is channel dependent, and has been detailed in section 2.
For each channel, the second mass cut for which we
tabulate results is the minimum for which we deem the
the EWA/ET approximation to be reliable
for all the different models.
For instance, in the case of the SM, Scalar and $O(2N)$ models, the optimal cut
on $\mll$ in the $\wp\wm$ channel is of order $500\gev$.  For such
a cut, contributing $\mww$ values are
large enough that the EWA/ET approximation is quite good.
Results for other cuts illustrate how rapidly the event levels
fall off with increasing invariant mass cut.
Lower cuts might be reliable for some channels in the case of
some models.  For instance, in the $\wp\wm$ channel,
the LET CG, LET K, and Delay K models all lack resonance structure
of any kind, and the EWA/ET approximation might be
adequate for $\mll\gsim 250\gev$.
In Fig.3 we show distributions in the mass variables for
several different models at the SSC.  Of course, the number of events
expected (see the tables) is generally much too small to allow
for an actual measurement of these mass distributions.  However,
the distributions allow for some intuitive feeling as to
where the signal event rates are largest and how rapidly the rates
decline with increasing invariant mass.

{}From the tables it is apparent that
the absolute number of (leptonic channel) signal events
in one SSC $10\fbi$ year or one LHC $100\fbi$ year is never large.
However, our cuts have reduced backgrounds to a remarkably low level,
so that even a relatively small number of excess $LL $ events should
be observable. Consequently, we find that for each model,
whether it has a scalar resonance, a vector resonance, or no resonance at all,
there is always a $WW$ charged-lepton mode for which the signal
is larger than background.
For instance, for the SM, the Scalar resonance model and the
$O(2N)$ model, the electroweak symmetry sector
contains a spin--zero isospin--zero resonance. As a result, the signal rates
in the $ZZ$ and $\wp\wm$ channels are
clearly above the background. Similarly,
the signal event rate in the $W^+Z$ mode is larger than the
background rate for the two Vector models
with a spin--one isospin--one resonance. Finally,
the $W^+W^+$ mode has the most significant
event rate in the LET CG, LET K, and Delay K models that have no resonance.

To quantify the observability of a given signal above
background, we proceed as follows. We define the signal to be observable
at a confidence level of $P\%$ if the maximum number of
background events, $\bmax$ at $P\%$ confidence
level is smaller than the minimum
number of signal {\it plus} background events, $\sbmin$,
at $P\%$ confidence level. Here, $\bmax$ is the number
of background events such that the probability of having
any number up to and including $\bmax$ is $0.01\,P$, while
$\sbmin$ is the number of signal plus background events such
that the probability of having a number greater than or equal to $\sbmin$
is $0.01\,P$. For a 99\% confidence level signal, these
two probabilities are 0.99.
The values $\bmax$ and $\sbmin$ are computed assuming Poisson statistics.
For integer values, this means that we require
$$
0.01\,P=\sum_{n=0}^{\bmax} {B^n\over n!}e^{-B};\quad
0.01\,P=\sum_{n=\sbmin}^{\infty}{(S+B)^n\over n!}e^{-(S+B)}\,,
\eqn\bsbdef$$
where $B$ and $S+B$ are the background and signal plus background
rates, respectively.
\foot{If the equalities in Eq.~\bsbdef\ are not satisfied for integer values
of $\bmax$ and $\sbmin$, we determine $\bmax$ and/or $\sbmin$ by
interpolating between the two integer values such
that the appropriate sum is just below and just above $0.01\,P$.}
$B$ and $S$ are obtained as a function of luminosity
from the event rates tabulated
earlier for the different types of models by scaling with
respect to the luminosities  of $10\fbi$ and $100\fbi$
adopted for the SSC and LHC, respectively, in constructing the tables.
We will uniformly employ results for the
middle (second) mass cut tabulated for each channel.
Even though backgrounds decrease rapidly with increasing
invariant mass cut, the signal also decreases (though less rapidly)
and the limited resulting statistics are such that
there is no channel for which the higher (third) mass cut tabulated in
Tables \ssclhctable a and \ssclhctable b leads
to a more observable signal at either the 99\% or 95\% confidence level.

As an example, consider again the $\wp\wm$ channel and the
SM, Scalar and $O(2N)$ models (for integrated luminosity of $10\fbi$
at the SSC). For all three models,
the background rate of 17 events is smaller than
the signal rates.  The smallest signal rate among the three models
for the $M_{\ell\ell}\geq 0.5\tev$ cut is
the 23 events predicted for the $O(2N)$ model.
The 99\% confidence level upper limit on the background is
27 events, whereas for the $O(2N)$ model the 99\% confidence level lower limit
on signal plus background is 25 events.
Thus, the predicted signal is not quite observable at the 99\%
confidence level for the $O(2N)$ model.  In contrast,
99\% confidence level {\it is} achieved for the SM and Scalar
models in the $\wp\wm$ channel after (less than) one $10\fbi$ year.

\TABLE\ssclum{}
\topinsert
\titlestyle{\twelvepoint
Table \ssclum a: Number of years (if $<10$) at SSC required for a
99\% confidence level signal.}
\bigskip
\hrule \vskip .04in \hrule
\thicksize=0pt
\begintable
Channel\\Model | SM & Scalar & $O(2N)$ & Vec. 2.0 & Vec. 2.5 & LET CG &
LET K & Delay K\cr
\qquad $ZZ$ \hfill    |
2.2  & 4.0    & 5.8  & \     & \     & 7.8  & \     & \    \nr
\qquad $\wp\wm$ \hfill |
0.50   & 1.0    & 1.2   & 2.5  & 3.5   & 2.5   & 4.0     & 4.0    \nr
\qquad $\wp Z$  \hfill |
\    & \    & \     & 1.5   & 2.8  & 3.2  & 4.2  & 2.8 \nr
\qquad $\wp\wp$ \hfill |
6.2  & 4.0    & 4.5   & 4.8  & 2.2  & 0.50   & 0.75  & 1.2 \endtable
\hrule \vskip .04in \hrule
\bigskip
\titlestyle{\twelvepoint
Table \ssclum b: Number of years (if $<10$) at SSC required for a
95\% confidence level signal.}
\bigskip
\hrule \vskip .04in \hrule
\begintable
Channel\\Model | SM & Scalar & $O(2N)$ & Vec. 2.0 & Vec. 2.5 & LET CG &
LET K & Delay K\cr
\qquad $ZZ$  \hfill   |
1.2  & 2.2 & 3.0     & \     & \     & 4.0     & 5.5   & \    \nr
\qquad $\wp\wm$ \hfill |
0.25  & 0.50  & 0.75  & 1.2  & 1.8  & 1.2  & 2.0     & 2.0    \nr
\qquad $\wp Z$  \hfill |
\    & \    & \     & 0.75  & 1.5   & 1.8  & 2.2  & 1.5  \nr
\qquad $\wp\wp$ \hfill |
3.2  & 2.2 & 2.2  & 2.5   & 1.2  & 0.25  & 0.50   & 0.50  \endtable
\hrule \vskip .04in \hrule
\endinsert

\TABLE\lhclum{}
\topinsert
\titlestyle{\twelvepoint
Table \lhclum a: Number of years (if $<10$) at LHC required for a
99\% confidence level signal.}
\bigskip
\hrule \vskip .04in \hrule
\thicksize=0pt
\begintable
Channel\\Model | SM & Scalar & $O(2N)$ & Vec. 2.0 & Vec. 2.5 & LET CG &
LET K & Delay K\cr
\qquad $ZZ$ \hfill     |
2.0     & 3.0    & 4.8  & \     & \     & 9.0     & \     & \    \nr
\qquad $\wp\wm$ \hfill |
0.75  & 1.2    & 2.0   & 7.5 & \ & 6.0 & \ & \    \nr
\qquad $\wp Z$ \hfill  |
 \    & \    & \     & 3.0     & 6.8  & 7.8  & 9.5   & 7.2 \nr
\qquad $\wp\wp$ \hfill |
5.2  & 3.2 & 4.2  & 3.5   & 2.0     & 0.75  & 0.75  & 1.8 \endtable
\hrule \vskip .04in \hrule
\bigskip
\titlestyle{\twelvepoint
Table \lhclum b: Number of years (if $<10$) at LHC required for a
95\% confidence level signal.}
\bigskip
\hrule \vskip .04in \hrule
\begintable
Channel\\Model | SM & Scalar & $O(2N)$ & Vec. 2.0 & Vec. 2.5 & LET CG &
LET K & Delay K\cr
\qquad $ZZ$ \hfill     |
1.2  & 1.8 & 2.5   & \     & \     & 4.8  & 6.0     & \    \nr
\qquad $\wp\wm$ \hfill |
0.50   & 0.75  & 1.0  & 3.8   & 5.2 & 3.0 & 5.0 & 6.5 \nr
\qquad $\wp Z$ \hfill  |
 \    & \    & \     & 1.8  & 3.5   & 4.0     & 4.8  & 3.8 \nr
\qquad $\wp\wp$ \hfill |
2.8  & 1.8 & 2.2  & 2.0     & 1.0     & 0.50   & 0.50   & 1.0    \endtable
\hrule \vskip .04in \hrule
\endinsert

The simplest manner in which the observability of all the various
signals can be tabulated is to give the number of years
required to achieve a signal at a given confidence level
for each channel and each model.  (Of course, if the machine can
be run at a higher instantaneous luminosity the required integrated
luminosities
can be achieved in less than the time indicated.)
These results for the SSC and LHC appear in Tables
\ssclum\ and \lhclum, respectively.
Let us first discuss the SSC results. In Table \ssclum a (\ssclum b) we give
the number of $10\fbi$ years required to see a signal
in a given channel for a given model at a 99\% (95\%) confidence
level, as defined above. As indicated earlier, the clearest signals
for the SM, Scalar and $O(2N)$ models are obtained
in the $\wp\wm$ channel.  The Vector models are most
easily probed in the $\wp Z$ channel. (Note, however, that for
the Vector 2.5 model, the $\wp\wp$ channel is actually superior
after imposing the kinematic cuts listed in Table~1.)
Finally, the LET CG, LET K, and
Delay K models are only readily probed using the $\wp\wp$ channel.
This is not to say that other channels are useless, especially
if a 95\% confidence level signal is deemed adequate.
Relatively small numbers of years ($<2.5$) are required to observe a signal
at 95\% confidence level in all cases except: the $\wp Z$ channel
for the SM, Scalar and $O(2N)$ models; the $ZZ$ channel for the
$O(2N)$, Vector, and LET/Delay models; and the $\wp\wp$
channel for the SM.

\TABLE\sscphitable{}
\topinsert
\titlestyle{\twelvepoint
Table \sscphitable: Percentage decrease in the SSC event rate
for the $\wp\wm$ channel relative to the $\cosphill<-0.8$ cut results
presented in Table \ssclhctable a.}
\bigskip
\thicksize=0pt
\hrule \vskip .04in \hrule
\begintable
$W^+ W^-$ | Bkgd. & SM & Scalar & $O(2N)$ & Vec 2.0 & Vec 2.5 & LET
CG &    LET K & Delay K  \cr
$\cosphill<-0.96$ | 25 & 3.7 & 4.3 & 3.5 & 2.3 & 2.3 & 2.0 &
2.2 & 2.2 \endtable
\hrule \vskip .04in \hrule

\endinsert

Of course, in the case of the $\wp\wp$
and $\wp Z$ channels we may also add in the opposite sign modes.
Because of the fact that the down-quark distribution function is
smaller than that for the up quark at moderate-to-large $x$
(high invariant masses probe fairly sizeable $x$ values),
these event rates are always smaller than those of the positive charge
channels.  One finds that the $W^-W^-$ signal event rate is about $1/3
\sim 1/2$ of the $W^+W^+$ rate. Similarly,
the ratio of $W^-Z$ to $W^+Z$ signal event rates
is about $1/2\sim 2/3$ for the models considered here.
Meanwhile, the irreducible $TT+LT$ background rates decrease by about a factor
of 2/3 in both channels. By combining the channels of both charge, the
observability of the $\wp\wp+\wm\wm$ and $\wp Z+\wm Z$ signals
is somewhat enhanced over the results given in the tables.
 It is important to note that one of the best means for
checking that we are observing the signal of interest is
to measure the ratio of $\wp\wp$ to $\wm\wm$ and
$\wp Z$ to $\wm Z$, respectively.  Should the
ability of the detectors to discriminate between lepton
charges at high momentum be inadequate, we would expect
these ratios to be near unity.

In obtaining our LHC results, we have used
cuts (as detailed in section 2) that are closely
analogous to those employed for the SSC.  In so doing, we did
not attempt to optimize the cuts to the same extent as we did
for the SSC. Thus, it is possible that
the signal/background ratios could be improved, although
we do not anticipate that further
optimization would lead to any dramatic changes.
{}From Tables \lhclum a and \lhclum b
we see that in the $\wp Z$, $ZZ$ and $\wp\wp$ channels
the LHC with a 100~${\rm fb}^{-1}$ luminosity is roughly
equivalent to the SSC with $10 {\rm fb}^{-1}$, except for the Vector models.
Because of the large resonance masses, the Vector models
are much more difficult to see at the LHC than at the SSC.
In the $\wp\wm$ channel, the SSC has a distinct advantage over the LHC
for all models.  This is due to the relatively greater difficulty
in removing the $t\anti t$ background at the LHC.
Another issue of concern for the LHC is the large probability
of having multiple interactions in one crossing, yielding
many minimum-bias and mini-jet events superimposed on
each $WW$ event of interest.
This type of pileup is likely to significantly
increase the background levels beyond those computed here on
the basis of one collision per crossing.  In addition, isolation criteria,
the central jet-veto, and jet-tagging, all of which are central
to our analysis, could become much more difficult to implement.
In this case, detection of strong interactions in the various $W_LW_L$
channels would be substantially more difficult at the LHC
than at the SSC.

\chapter{Discussion}

\REF\wzc{M. Chanowitz, talk presented in the XXVI International Conf.
on High Energy Physics, Dallas, Aug., 1992;
and talk presented in the Workshop on Electroweak Symmetry Breaking
at Colliding-Beam Facilities, University of California at Santa Cruz,
December, 1992.}
\REF\wzdh{A.~Dobado, M.~J.~Herrero, and T.~N.~Truong, Phys.\ Lett.\ {\bf 235},
129 (1990).}
\REF\zznunu{R. N. Cahn and M. Chanowitz, Phys. Rev. Lett. {\bf 56}, 1327
(1986); V. Barger, T. Han, and R. Phillips, Phys. Rev. {\bf D36}, 295 (1987).}
\REF\jggkseidenmult{J.F. Gunion {\it et al.}, Phys. Rev. {\bf D40}, 2223
(1989). }
\REF\bjrap{J.~D.~Bjorken, Phys. Rev. {\bf D47}, 101 (1993);
SLAC-PUB-5823, talk given at the IXth International
Workshop on Photon-Photon Collisions, University of California, San Diego,
La Jolla, CA, March 23-27, 1992.}

An important question is the extent to which we have truly optimized
the procedures for isolating an $LL $ signal in the various $WW$
purely-leptonic final state channels. Below we discuss several
improvements that might turn out to be feasible.

A possible improvement in the significance of
the signals in the $\wp\wm$ and $\wp\wp$ channels can be obtained
by tightening the cut on $\cosphill$.  The improvement that
can be obtained, based on our EWA/ET
parton-level Monte Carlo, is illustrated in Table \sscphitable\ in the
case of the $\wp\wm$ channel. By tightening the cut from
$\cosphill<-0.8$ to $\cosphill<-0.96$, the background is reduced to 3/4
of its previous size
while the $LL $ signal rate is decreased by at most 4\%
(in the SM and Scalar cases) and perhaps, by as little as 2\%
(Vector and LET CG models). Such a large decrease in the background level
would clearly lead to an increase in the significance
of the signals in these channels.

However, these results were obtained using the EWA/ET calculation in which
the transverse momentum of the $\wl\wl$ pair, $p_T(WW)$, is ignored.
A non-zero value for $p_T(WW)$ would imply
that the $\wl\wl$ are not exactly back-to-back.
Previously, we noted the $1/(p_T^2+M_W^2)^2$ distribution of the
$\wl$'s which initiate the $\wl\wl$ scattering (as measured
with respect to the quarks from which they are emitted). This steep
fall-off implies that $p_T(WW)$ for the $\wl\wl$ signal is
very limited (in contrast to the $\wl\wt$ and $\wt\wt$ background).
Typically, $p_T(WW)$ for the $\wl\wl$ scattering signal
is not much larger than $M_W$.
For an event with $p_T(W) \sim 0.5$ TeV, $p_T(WW) \sim M_W$
can result in angle of $\phi\approx 162^\circ$
(for the configuration where $p_T(WW)$ is perpendicular to the
individual $p_T$'s of the $W$'s), which
corresponds to $\cos \phi (WW) \approx -0.95$.
Thus, although the agreement between the exact ``subtraction''
calculation and the EWA/ET calculation is excellent for a modest cut at
$-0.8$,
as illustrated by the $\wp\wp$ comparison of Table~\comptable,
this agreement might worsen if the cut is strengthened.  Indeed,
in the $\wp\wm$ channel we have found that the exact ``subtraction''
result with $\cosphill < -0.96$ is about 23\%
as much as that with
$\cosphill < -0.8$, as compared to the 4\% decrease
listed in Table \sscphitable\ obtained using the EWA/ET amplitudes.
Further, there are additional sources of $p_T(WW)$.
A Monte Carlo which goes beyond the
parton level will include initial-state radiation of gluons, the intrinsic
transverse momentum of the quarks that initially radiate the fusing $W$'s,
and hard gluon radiation in the final state (part of the higher--order
QCD corrections to the $W_LW_L$ scattering process).
All of these effects will tend to impart some net
transverse momentum to the $W_LW_L$ pair and decrease the
fraction of lepton pairs that are sufficiently back-to-back to pass
a very severe cut on $\cosphill$.  We anticipate that the $-0.8$
cut is sufficiently moderate that such effects will not significantly
alter the efficiencies obtained from a parton Monte Carlo.
It is worthwhile to notice that the effects of $p_T(WW)$
are less important
for models with more events in the larger mass region ($M(WW) > 1$~TeV),
such as LET CG. This was also demonstrated in Ref.~{\wpwpbc}, where an
empirical formula for the $p_T(WW)$ spectrum was used in combination with
the EWA.

Regardless of which $\cosphill$ cut turns out to be most appropriate,
one can ask whether it would be beneficial
in the $ZZ$ and $WZ$ channels. We
have not imposed this cut in our work
because background event rates in these channels
are already small after the cuts employed, and the amount of improvement
in the significance of a signal would be marginal.  It is only
if some of the cuts that we {\it have} employed must be significantly weakened,
or if our cuts are not so efficient in eliminating the background,
when the actual data is analyzed, that a $\cosphill$ cut in these channels
might prove valuable.

Another issue is the extent to which our cuts eliminate a contribution
to the $LL $ signals of interest arising from a source quite distinct
from the $W_LW_L$ scattering processes upon which we have focused.
An example of such a situation arises in the case of the $WZ$ channel.
If the appropriate model contains a spin--one isospin--one resonance,
then a larger signal rate is obtained by eliminating the jet-tag
cut.  This is because there is an additional contribution from
$q \anti q$ fusion in which a virtual $W$ is created that then
mixes with the vector resonance.
Some analysis of this situation has appeared in
Refs.~\refmark{\BESSssc,\wzc,\wzdh}, where it was
found that one might be able to observe a signal
without jet-tagging if such a resonance exists.
However, eliminating the jet-tagging is much more likely
to be viable at the SSC than at the LHC.  In Tables~\smssctable a
and \smssctable b,
we saw that the signal/background ratio becomes much worse at
the LHC than at the SSC if jet-tagging is not performed.
Clearly, a careful study is required; this is
beyond the scope of the present paper.

We must not forget that, for each channel,
we have considered only the ``gold--plated'' purely-leptonic decay modes
containing the maximum possible number of charged leptons.
These are the cleanest modes for observing the $LL $ signal, but
a significant price is paid in terms of branching ratios.
The next-most clean mode that can be considered is $ZZ\rta \lplm
\nu\anti\nu$. This mode has roughly six times as large a branching ratio
as the four-charged-lepton mode we have studied.
Parton-level calculations \refmark{\zznunu} and some  recent SDC
detector studies \refmark{\sdc}
indicate that cuts can be implemented
which could eliminate reducible backgrounds in this mode.
The only issue is the extent to which the irreducible
EW $Z_TZ_T+Z_TZ_L$ backgrounds can be suppressed.
Some study of this has appeared in Ref. \wzc.  There, it is found
that a significant improvement in the observability of the $Z_LZ_L$
signal can be obtained for several models if the $\lplm\nu\anti\nu$
mode is employed.

Of course, $WW$ final states containing a mixture of
leptons and jets have still higher branching ratios.
However, mixed QCD-electroweak backgrounds enter.
Many of the techniques that we have developed here for isolating the purely
leptonic signals will also be applicable for such mixed states,
and additional cuts will become relevant, \eg\ a cut on multiplicity
and/or rapidity structure \refmark{\jggkseidenmult,\bjrap}.
A refined study of the mixed modes, incorporating some of the procedures
that have been developed for the purely leptonic modes, should
be performed \refmark{\rwwww}, but is beyond the scope of this paper.

We have demonstrated the importance of using the single jet-tagging to enhance
the signal/background ratio, especially for the $W^+W^-$ mode to suppress the
huge $t \bar t$ background \refmark{\hww}.
Our resulting tagging efficiency for the signal
agrees well with the full Monte Carlo study for the SDC detector
\refmark{\sdc}.
This is also true for the background process $t \bar t j$;
\foot{In preliminary versions of this work, a programming error led
to an apparent disagreement \refmark{\hww}.}
our fixed order
$\alpha_s^3$ parton-level calculation agrees quite well with results quoted
in the SDC report \refmark{\sdc}. Nonetheless,
still more careful studies of jet-tagging in the forward/backward
region would be worthwhile.

\chapter{Conclusions}

We have demonstrated that, at both the SSC and the LHC,
viable signals for strong $W_LW_L$ interactions
can be obtained for a wide variety of models in the purely leptonic
final states.  Of course,
the channels examined, $\wp\wm\rta \lplm \nu\anti\nu$, $\wp Z\rta \lplm
\ell\nu$,
$ZZ\rta \lplm\lplm$, and $\wp\wp\rta \lplp\nu\nu$, do not {\it all} yield
adequate signals in 1-2 years of canonical SSC or LHC luminosity for
{\it all} models.
Instead, we find that a significant signal can always be found
in the channels that most naturally complement the particular type
of model considered.  In particular, models with a resonance of definite
isospin are most easily probed using the $WW$ channels that have resonant
contributions from that same isospin. Indeed, one of our more important
conclusions is that different types of models can be distinguished
experimentally by determining the relative magnitude of the $LL$ signals in
the four channels listed above.

A large part of our work focused on the techniques required to suppress
reducible and, especially, irreducible backgrounds to a level such
that the low $LL$ signal event rates in the purely leptonic channels
can be isolated. In particular, the irreducible backgrounds
from production of $WW$ pairs with $TT$ and $LT$ polarizations
end up being most important, and our techniques are particularly
focused on suppressing them. Although our calculations do not include detector
effects, we believe that they will survive more sophisticated Monte Carlo
analyses.  In particular, the types of cuts we have employed should be
directly applicable in the experimental analyses
that will be performed when actual data becomes available.

Overall, we conclude that it is possible to probe a strongly
interacting electroweak symmetry breaking sector at the SSC or LHC using only
the ``gold--plated'' purely-leptonic modes studied here.
Even if a light Higgs boson is found, it will be important
to measure the event rates at high $\mww$ in all the various channels
in order to make certain that the Higgs boson
completely cures the bad high-energy behavior in all $WW$ scattering
subprocesses. The low event rates for the purely-leptonic final states imply
that of order 2-3 years of $10\fbi$ annual luminosity
will be required to conclude that there is no obvious $W_LW_L$
enhancement in any of the four channels.
Because of the relative cleanliness of these final states, the option of
achieving this required integrated luminosity via enhanced instantaneous
luminosity should be strongly considered.

\ack
We would like to thank M.~Chanowitz, M.~Berger, S.~Dawson, G.~Kane, S.~Mrenna,
S.~Naculich, J.~Ohnemus, L.~Orr, F.~Paige, R.~Phillips,
C.~Quigg, W.~Repko, R.~Stuart,
G.~Valencia, E.~Wang, S.~Willenbrock, H.~Yamamoto,
and D.~Zeppenfeld for many helpful discussions.
The work of J. Bagger was supported in part by NSF grant
PHY-9096198 and in part by Texas National Research Laboratory
Commission (TNRLC) grant RGFY93292. The work of V.~Barger was supported
in part by DOE grant DE-AC02-76ER00881 and in part by TNRLC grant RGFY9273.
The work of K.~Cheung was supported in part by DOE grant DE-FG02-91-ER40684.
The work of J. Gunion was supported in part by DOE grant
DE-FG-03-91ER40674. T. Han was supported in part by TNRLC Awards No. FCFY9116.
The work of G. Ladinsky was supported in part by DOE grant
DE-FG02-90ER-40577 and TNRLC grant RGFY9240.
The work of R. Rosenfeld and C.--P. Yuan was supported in part by
TNRLC grant RGFY9114 and RGFY9240, respectively.

\endpage
\refout
\endpage
\figout
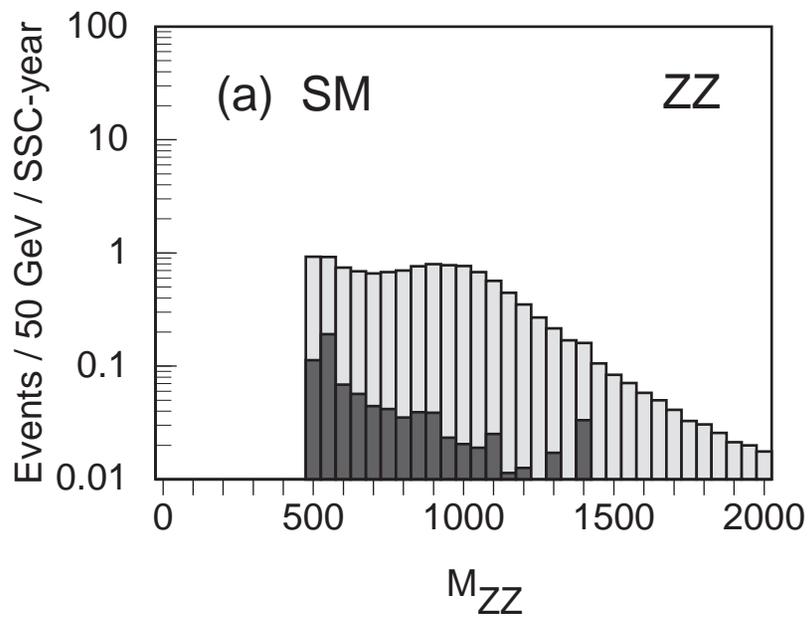
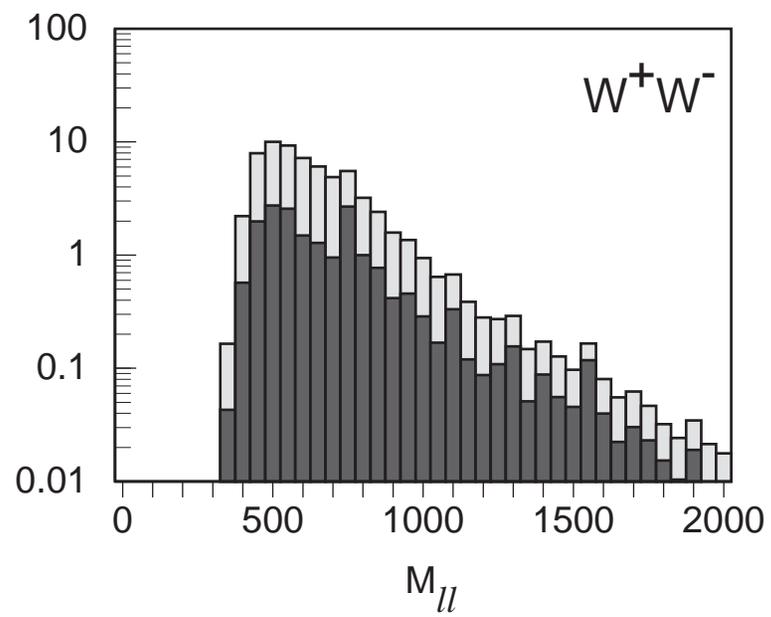
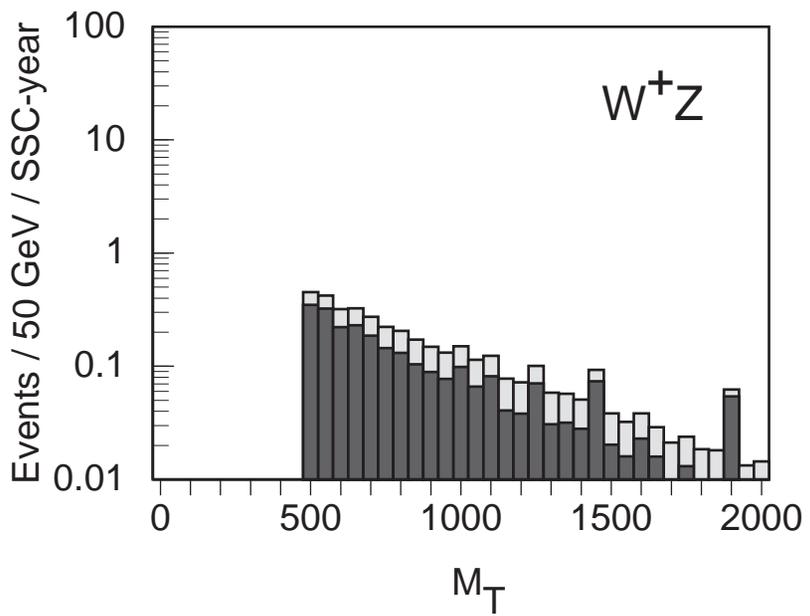
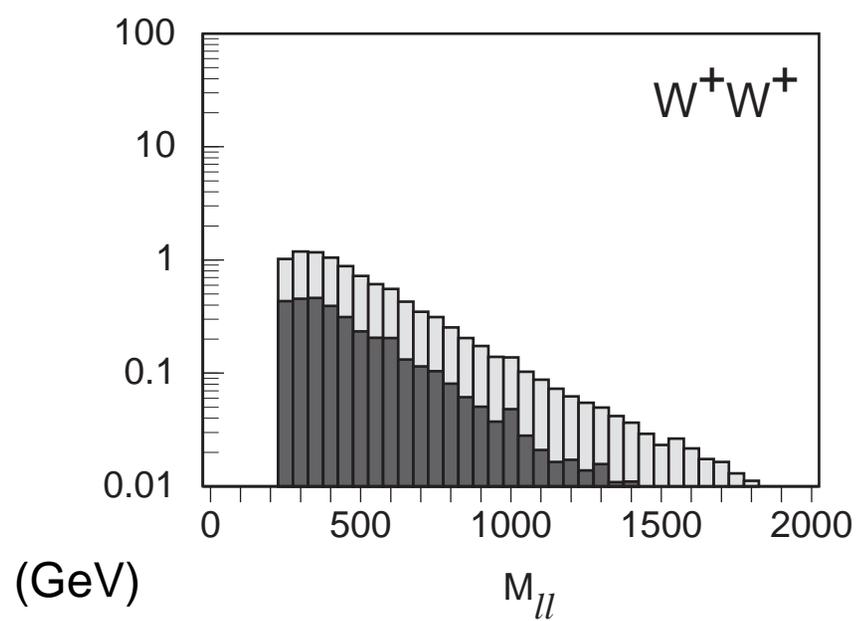
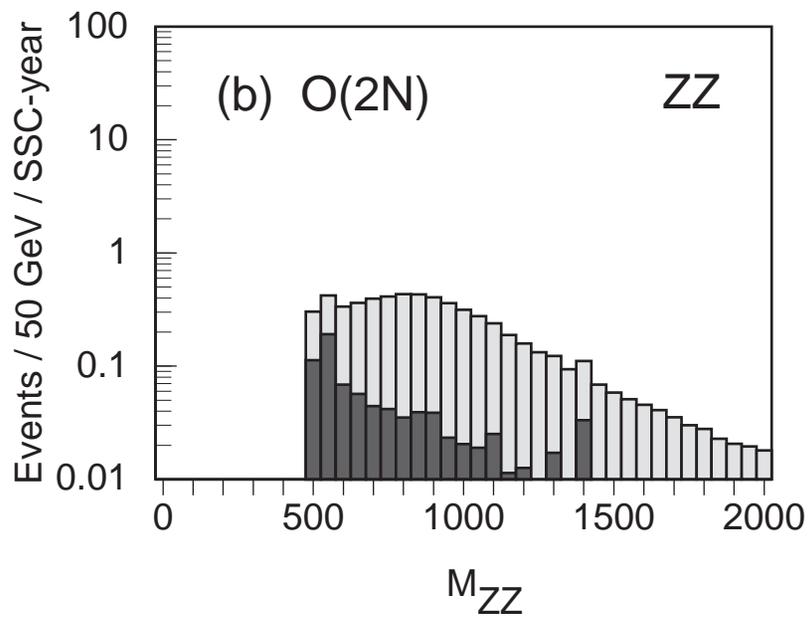
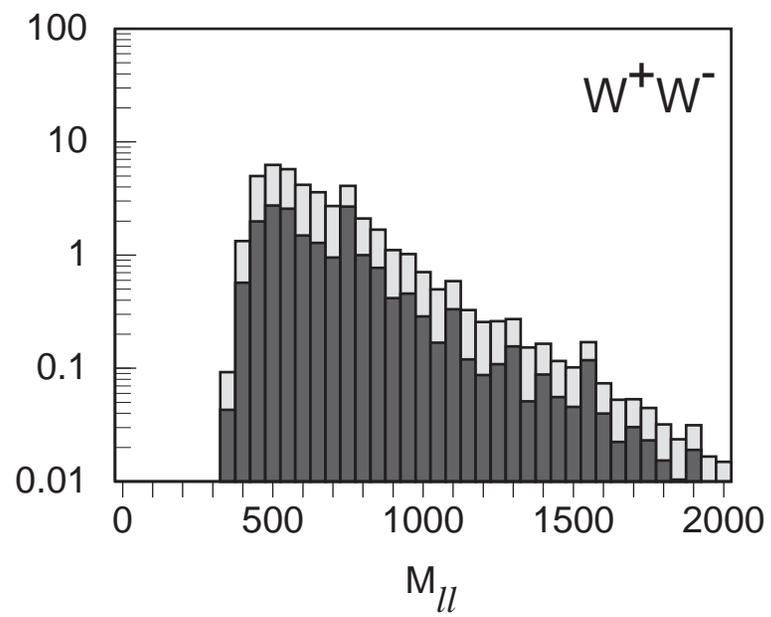
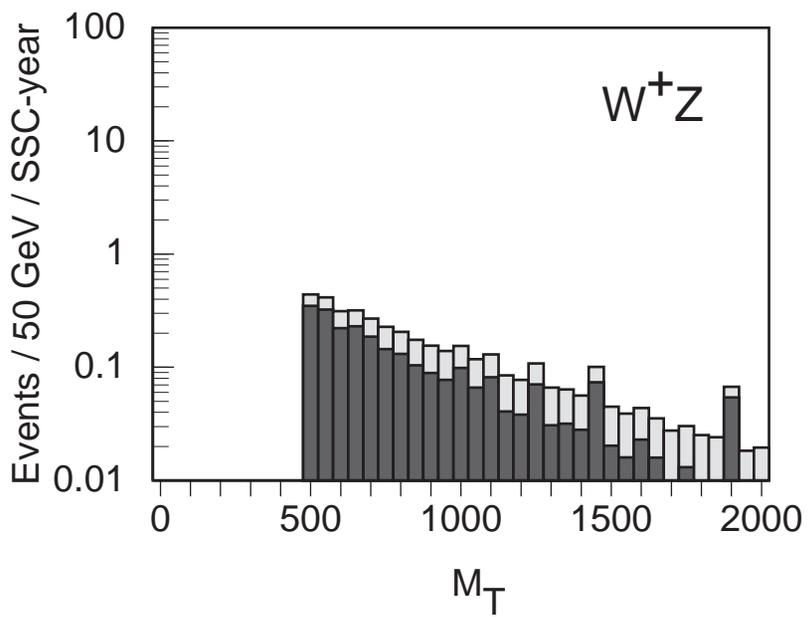
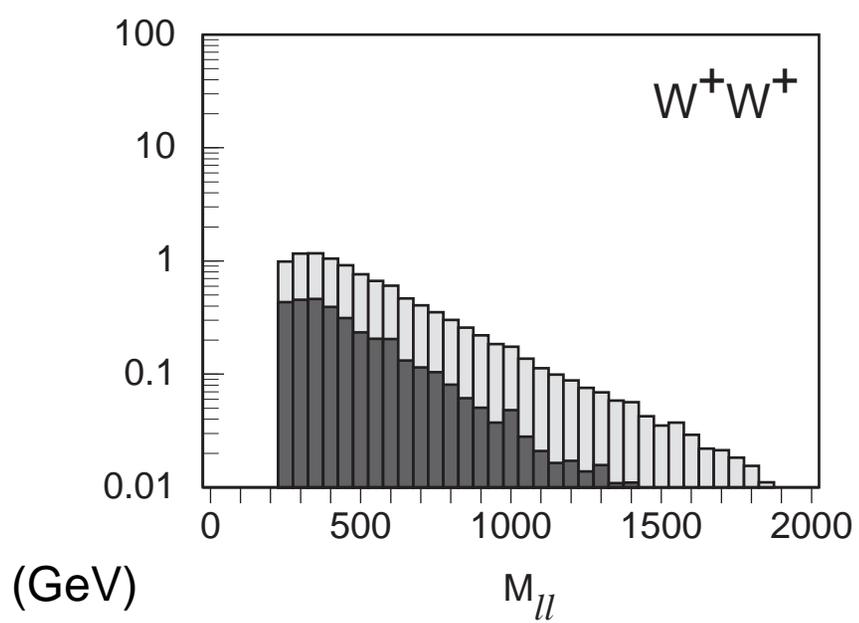
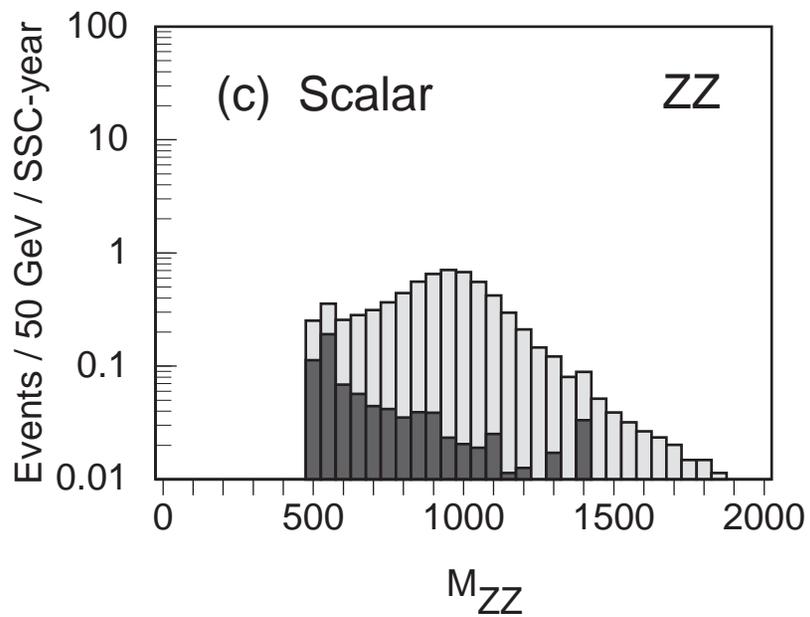
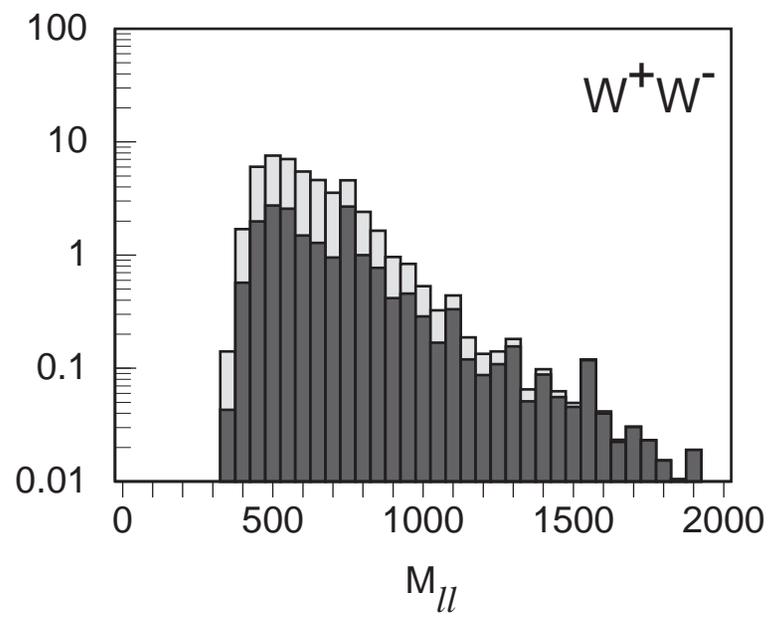
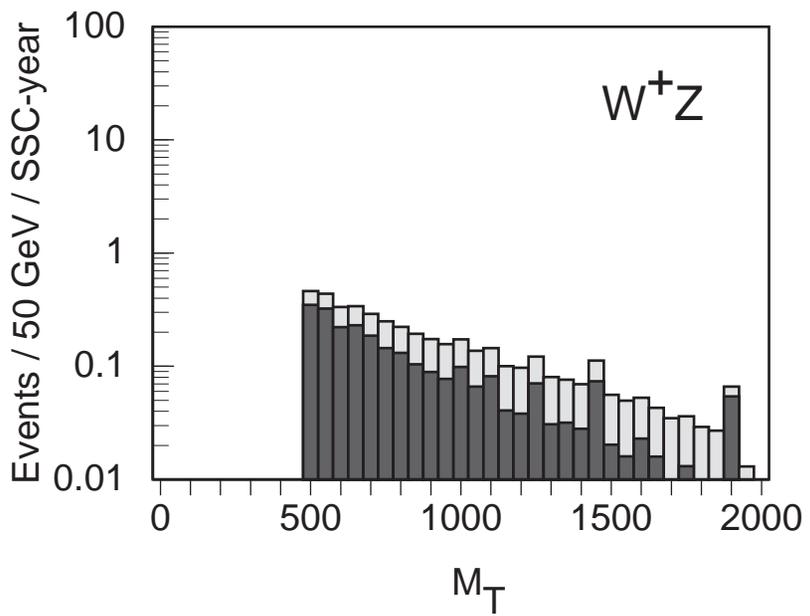
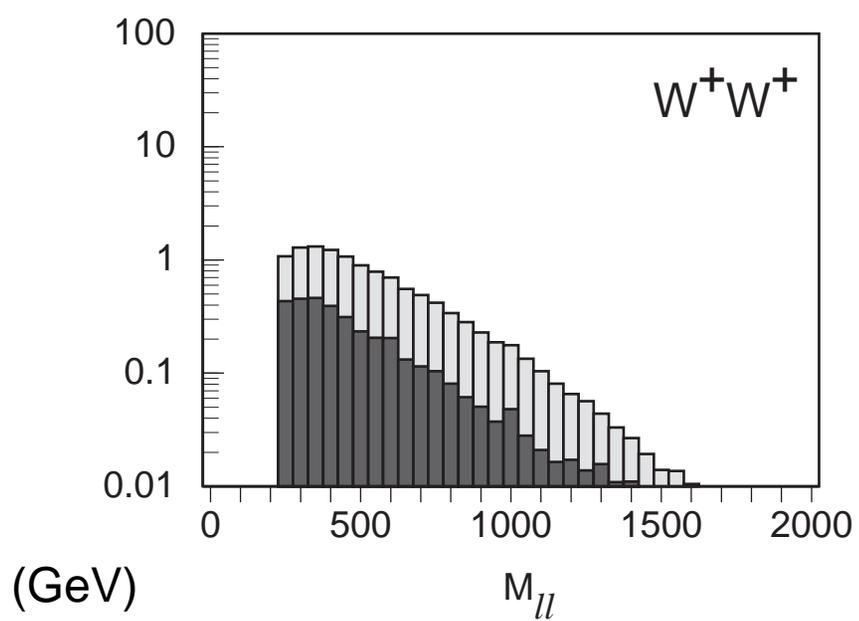
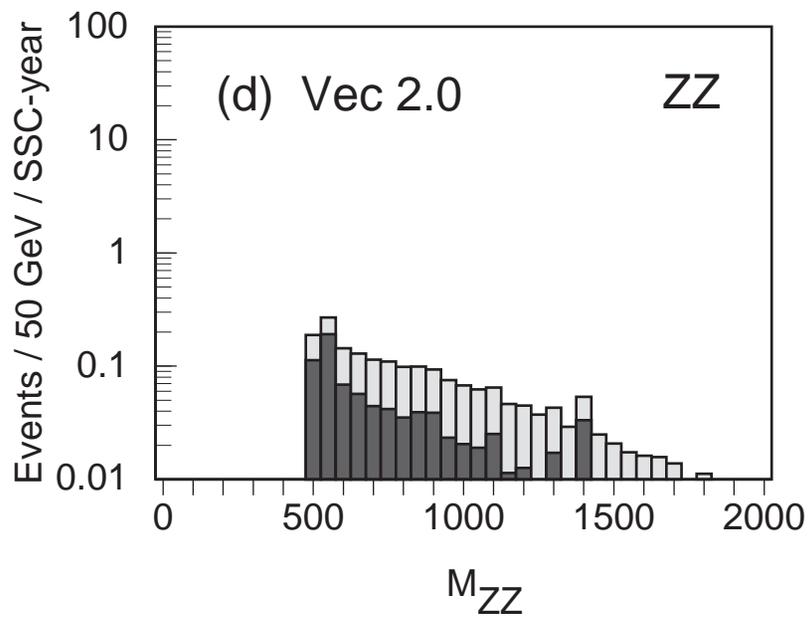
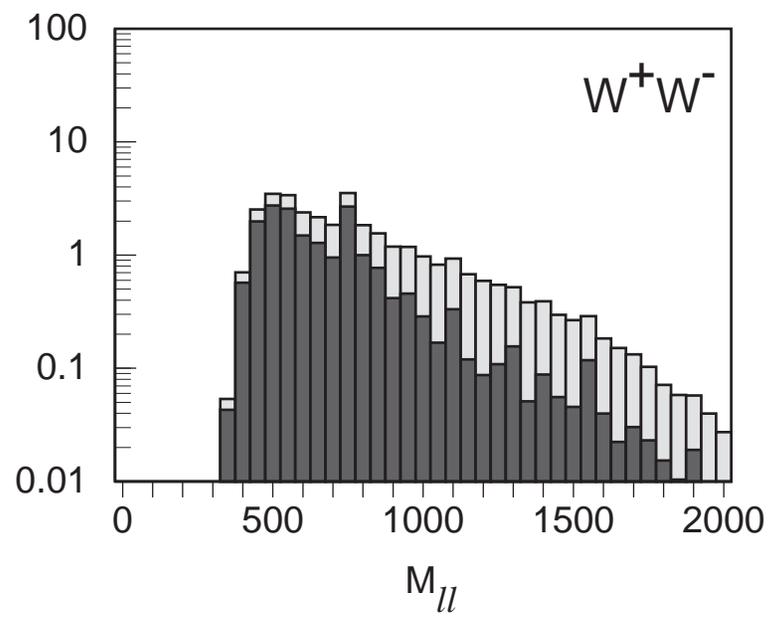
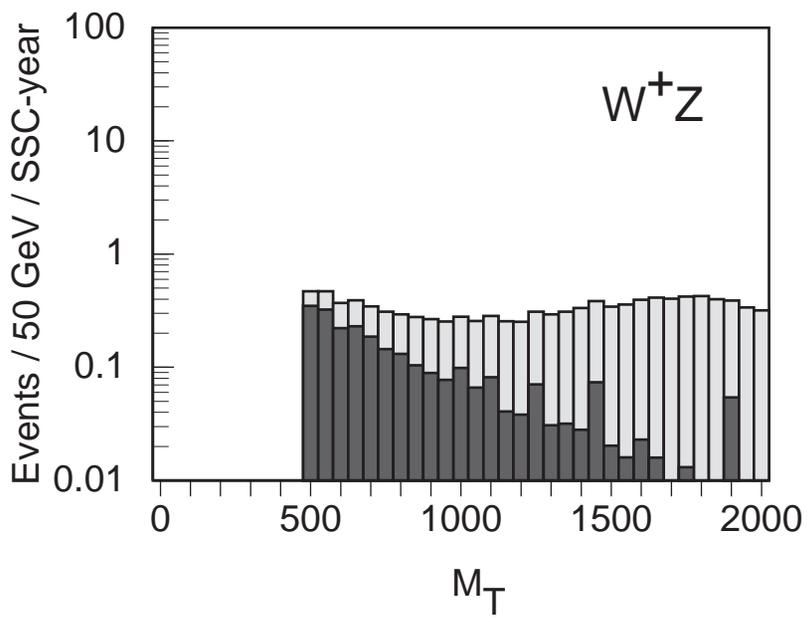
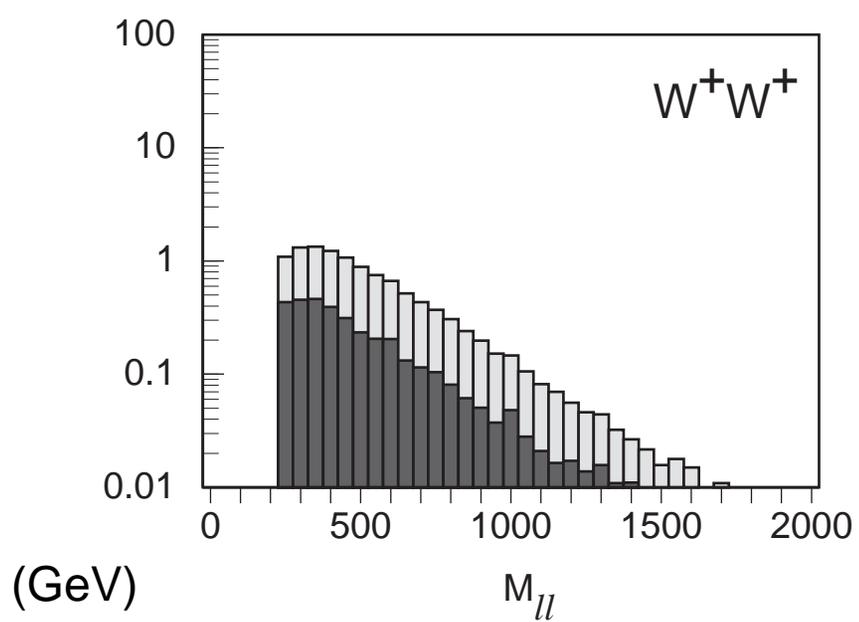
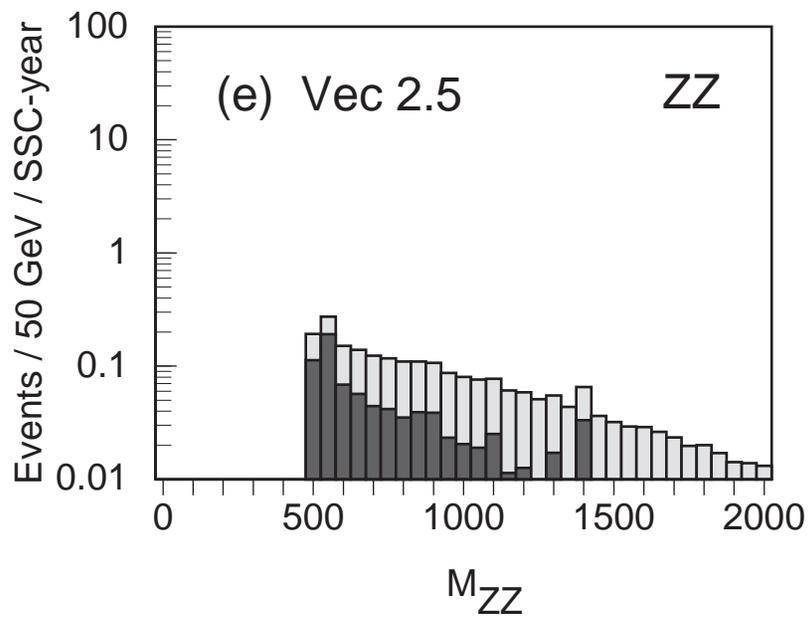
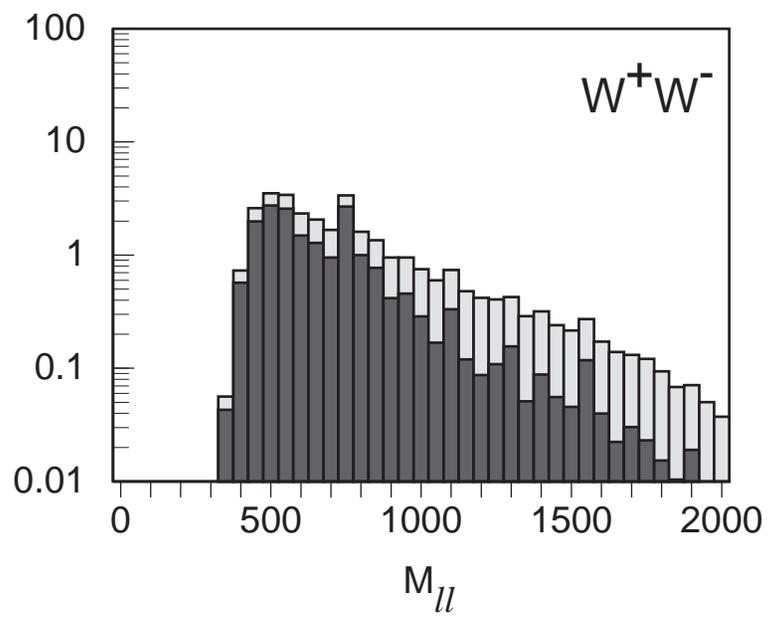
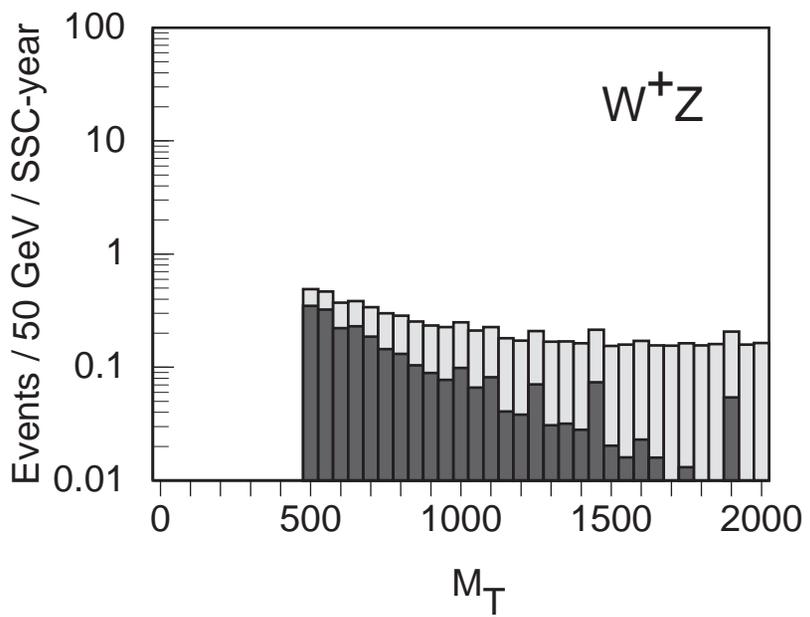
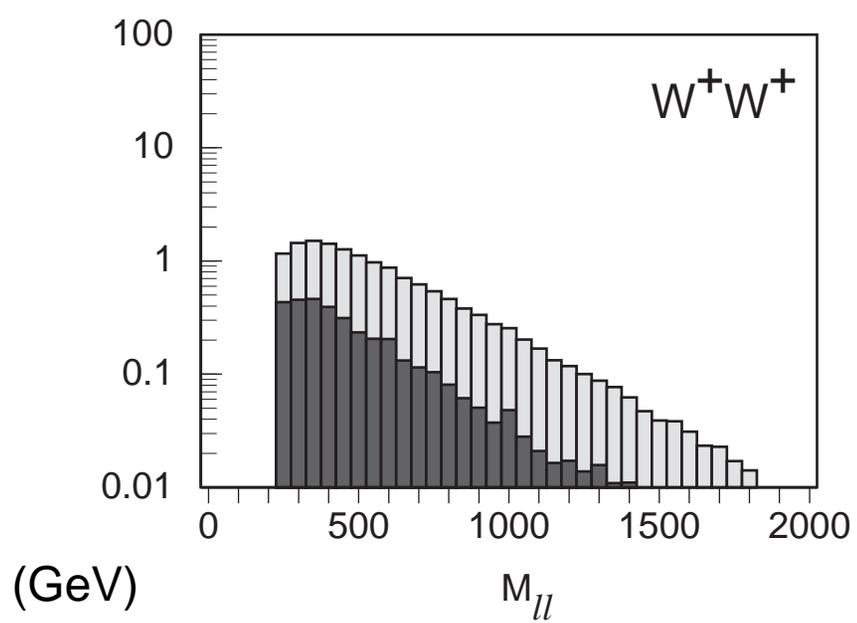
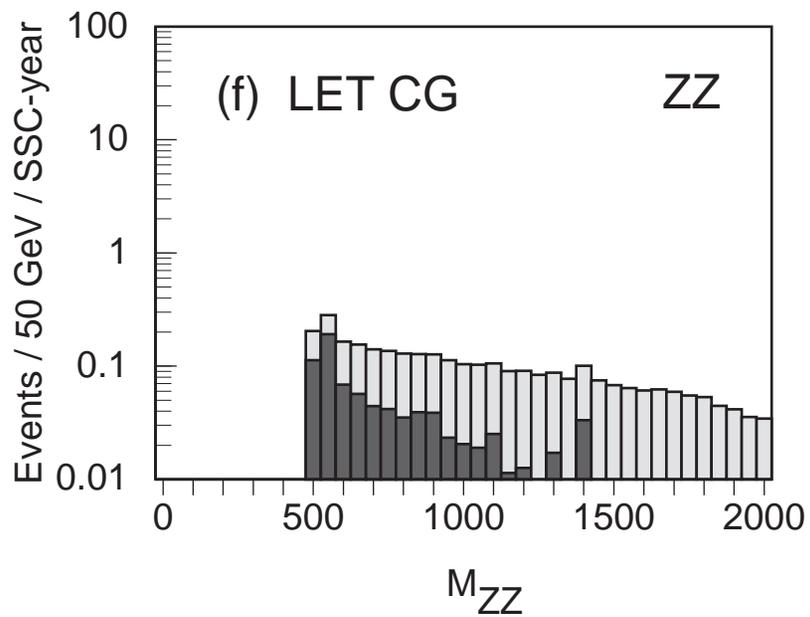
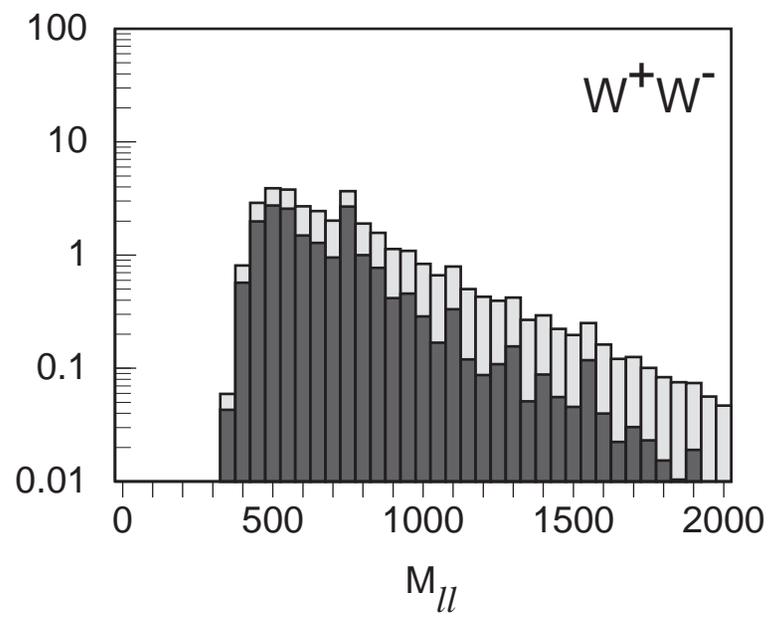
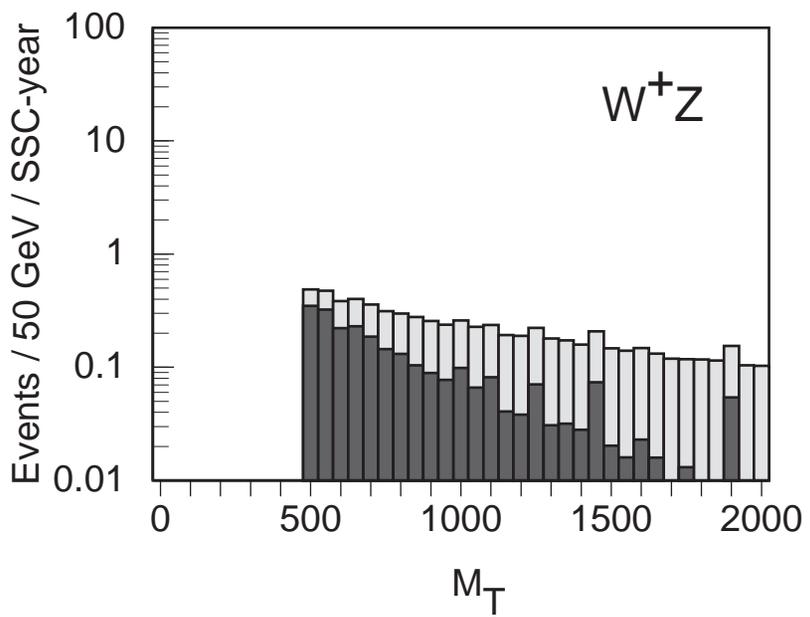
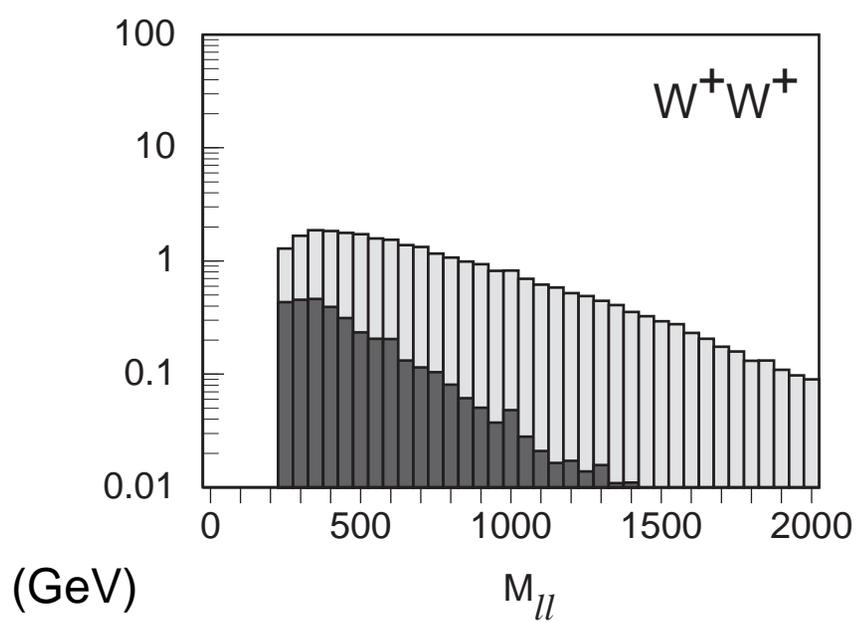
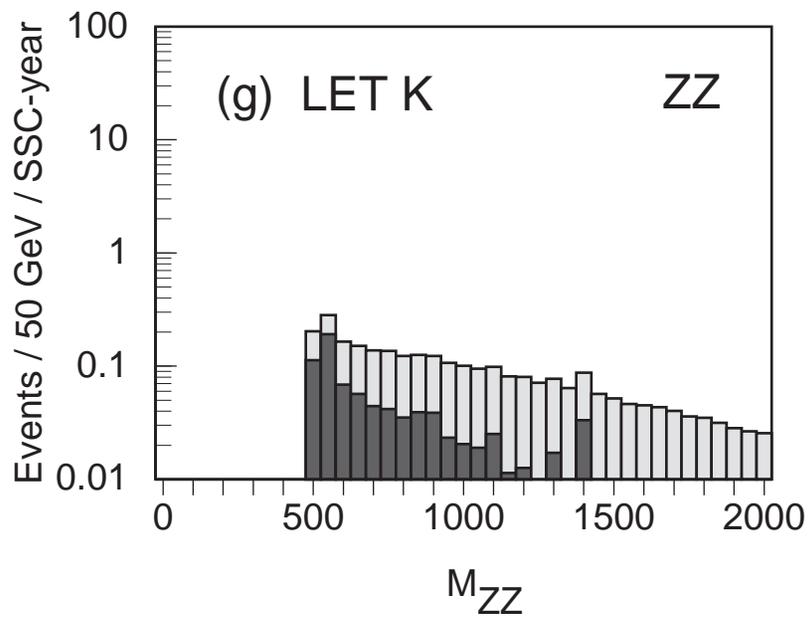
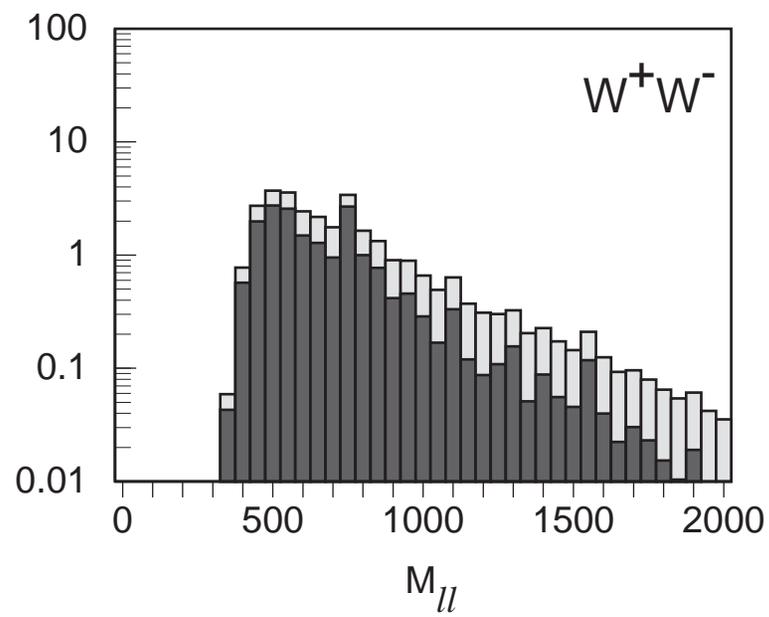
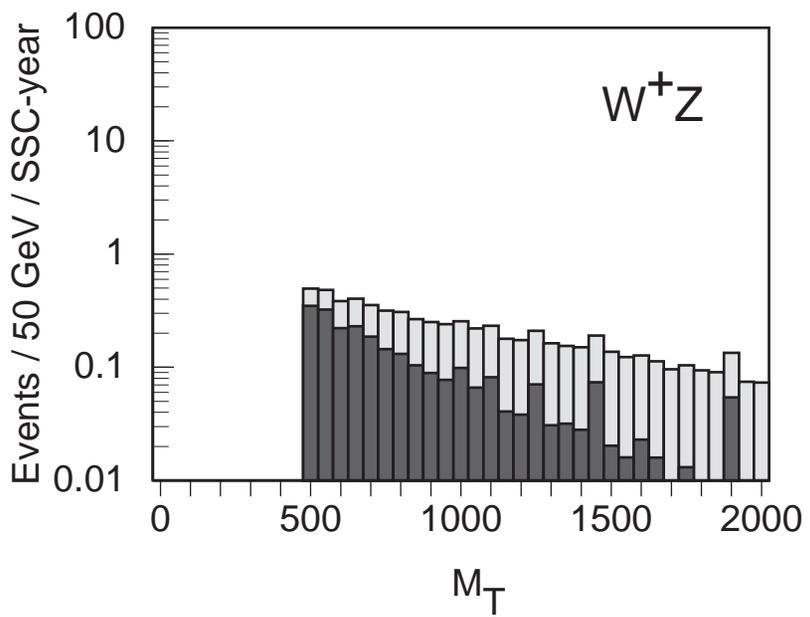
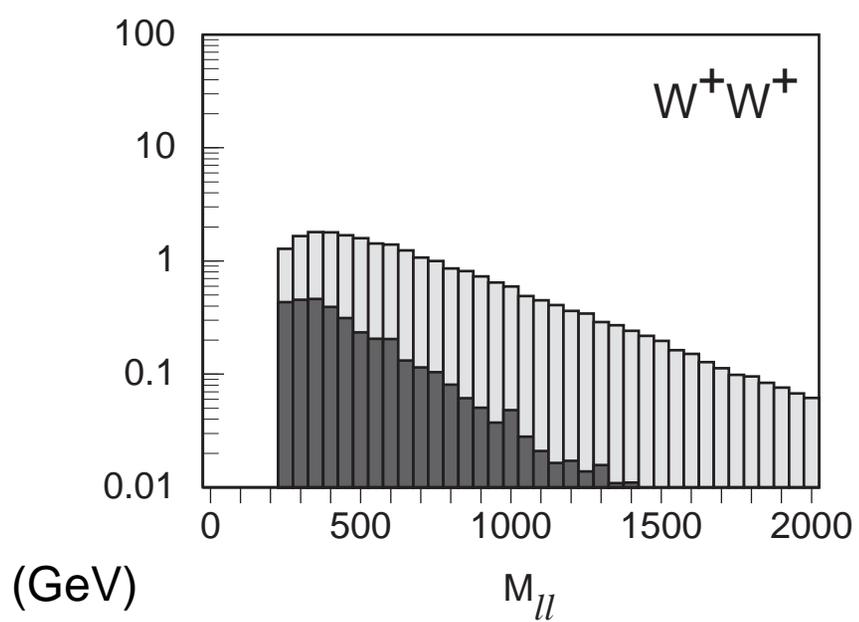
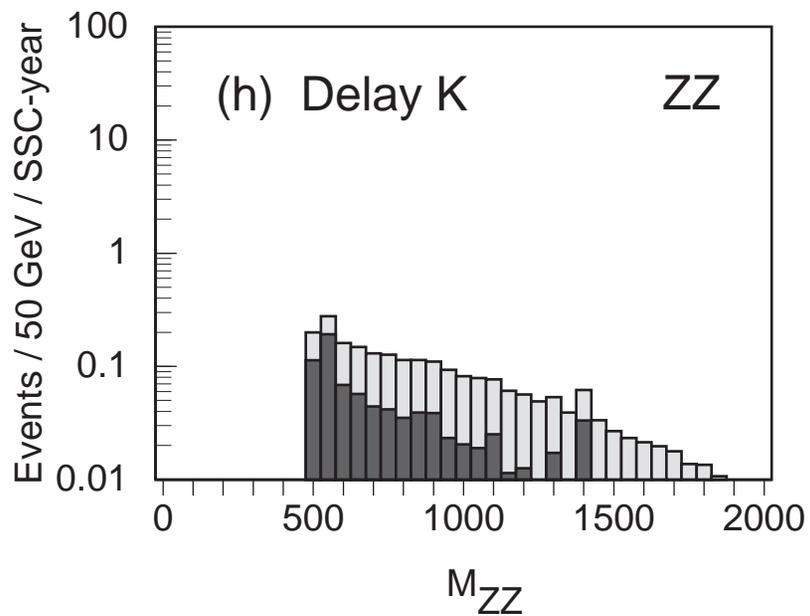
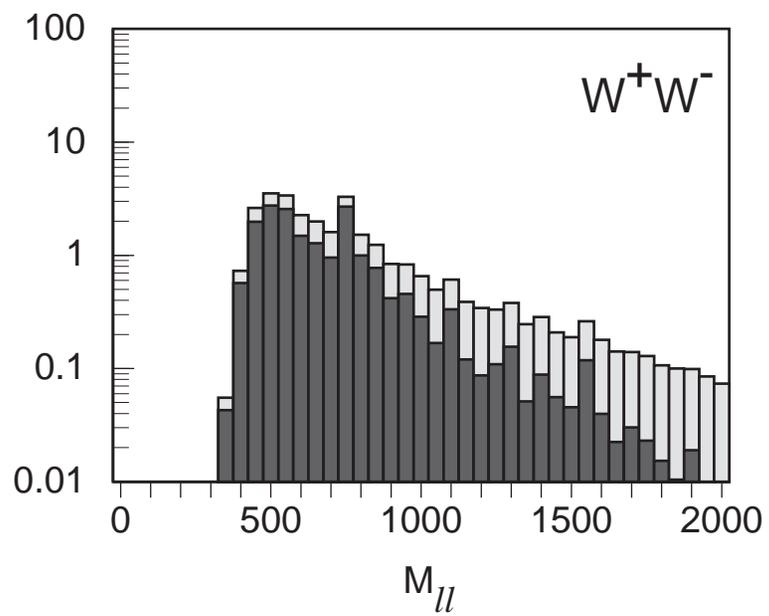
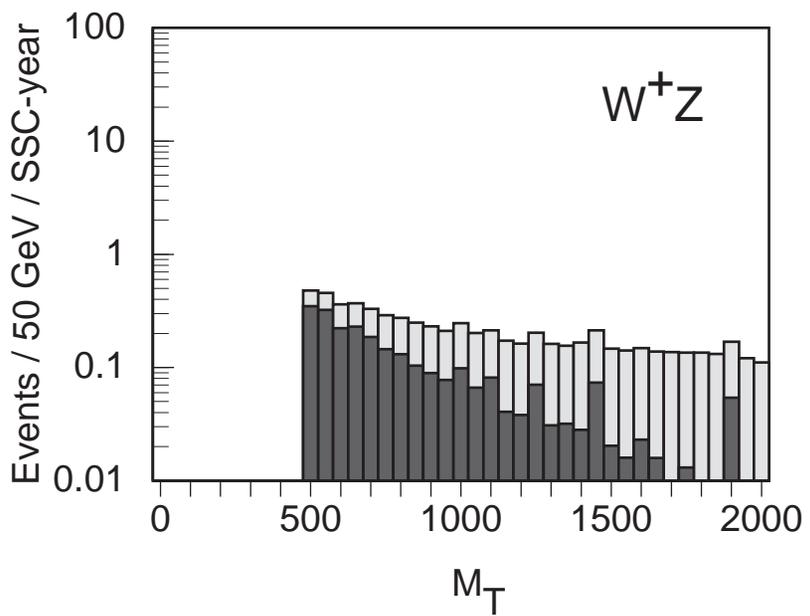
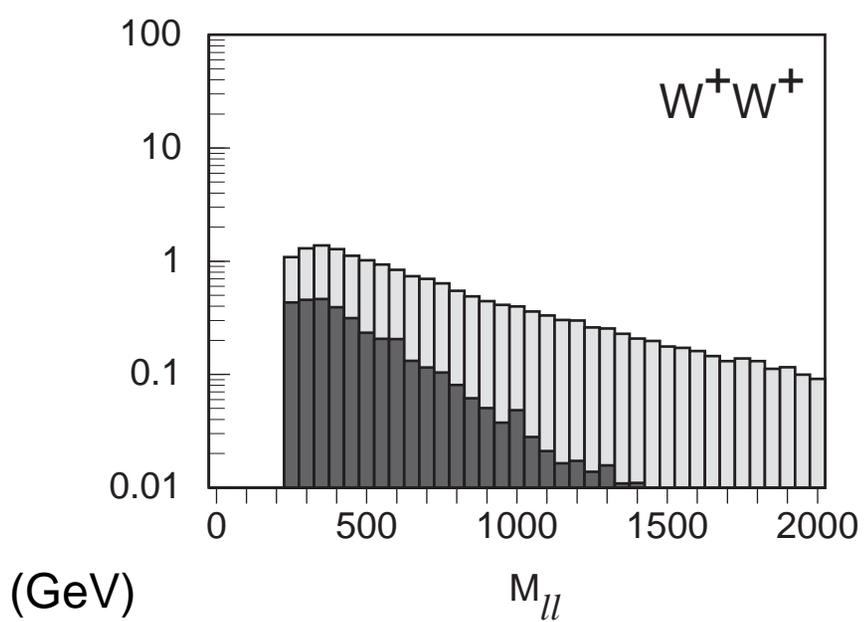
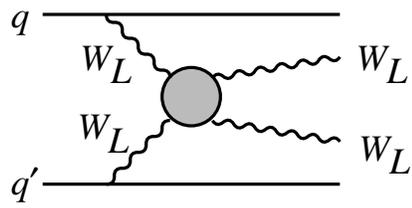
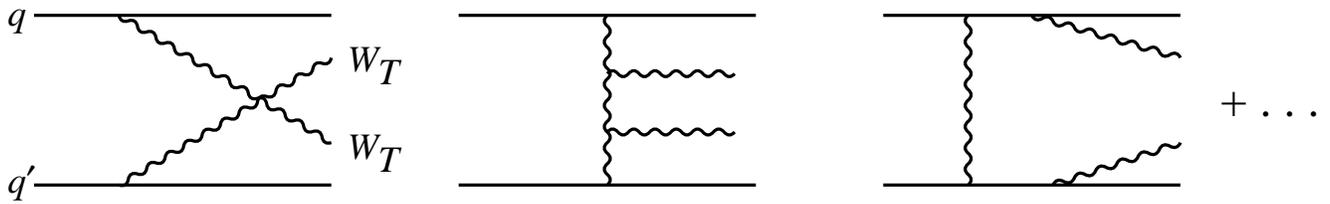
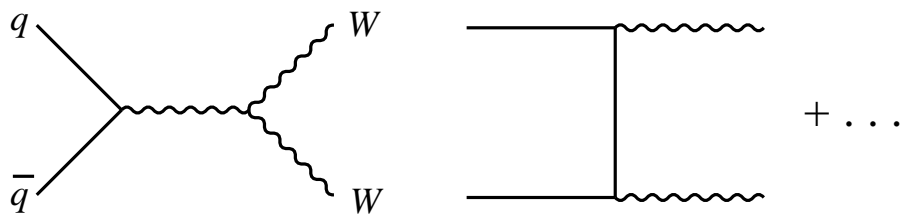
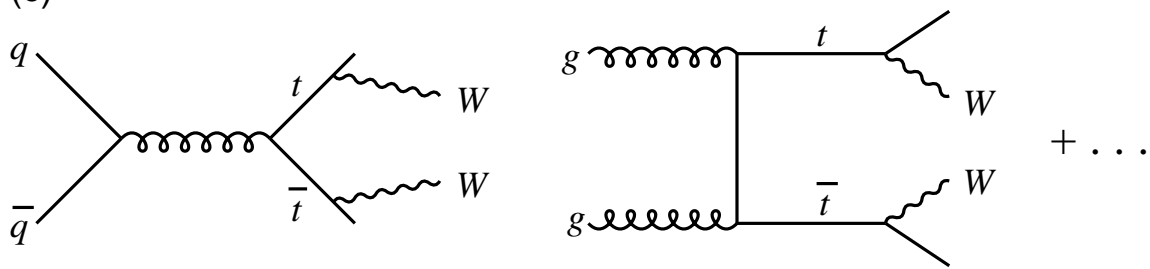
\end